\PassOptionsToPackage{nameinlink}{cleveref}
\documentclass[a4paper,cleveref, thm-restate]{lipics-v2021}

\hideLIPIcs  

\usepackage{xcolor}

\usepackage[T1]{fontenc}
\usepackage{mathtools}

\usepackage{xspace}

\RequirePackage{todonotes}

\usepackage[notion, quotation, electronic]{knowledge}
\definecolor{Blue Sapphire}{HTML}{005f73}
\definecolor{Blue Matt}{HTML}{094c7b}
\definecolor{Blue Dark}{HTML}{1B4C6E}
\definecolor{Golden}{HTML}{E27400}
\definecolor{Brown}{HTML}{ae740e}
\definecolor{Gamboge}{HTML}{ee9b00}
\definecolor{Ruby Red}{HTML}{9b2226}
\definecolor{Blue Marine}{HTML}{022687}
\definecolor{Dark Ruby Red}{HTML}{580507}
\definecolor{Dark Blue Sapphire}{HTML}{053641}
\definecolor{Dark Gamboge}{HTML}{be7c00}
\definecolor{Blue Marine}{HTML}{022687}

\IfKnowledgePaperModeTF{
	}{
		\knowledgestyle{intro notion}{color={Dark Ruby Red}, emphasize}
		\knowledgestyle{notion}{color={Dark Blue Sapphire}}
		\hypersetup{
				colorlinks=true,
				breaklinks=true,
				linkcolor={}, 
				citecolor={}, 
				filecolor={Blue Marine}, 
				urlcolor={Blue Marine},
			}
	}

\knowledge{notion}
 | A
 | Conjecture A

\knowledge{notion}
 | Borel \mso

\knowledge{notion}
 | B
 | Conjecture B

\knowledge{notion}
 | Muller-Borel game
 | Muller-Borel games

\knowledge{notion}
 | finite memory
 | finite-memory strategy

\knowledge{notion}
 | nodes

\knowledge{notion}
 | branches
 | branching

\knowledge{notion}
 | Borel tree
 | Borel trees

\knowledge{notion}
 | nondeterministic Muller automaton on Borel trees
 | nondeterministic Muller automata on Borel trees
 | Muller automaton on Borel trees

\knowledge{notion}
 | game

\knowledge{notion}
 | arena

\knowledge{notion}
 | positions

\knowledge{notion}
 | strategy

\knowledge{notion}
 | histories
 | history

\knowledge{notion}
 | play

\knowledge{notion}
 | finite-memory strategies

\knowledge{notion}
 | prioritized Borel game
 | prioritized Borel games
 | game@Borel

\knowledge{notion}
 | Priority
 | priority
 | priorities

\knowledge{notion}
 | determined

\knowledge{notion}
 | winner

\knowledge{notion}
 | memoryless
 | memoryless strategy

\knowledge{notion}
 | open
 | open sets
 | topology

\knowledge{notion}
 | witnesses
 | witness

\knowledge{notion}
 | $(\imax+1)$-witnesses
 | $(k+1)$-witness
 | $m$-witness
 | $1$-witness
 | $\imax $-witnesses
 | $\dmax $-witness
 | $i$-witness
 | $i_{0}$-witnesses
 | $j$-witness
 | $i_0$-witness
 | $i_0$-witnesses
 | $i'$-witness
 | $\imax $-witness
 | $i$-witnesses
 | $j$-witnesses

\knowledge{notion}
 | profile@two
 | profiles@two

\knowledge{notion}
 | alternatingly nested

\knowledge{notion}
 | $p$-profile
 | $p$-profiles

\knowledge{notion}
 | Branching nodes
 | branching nodes
 | branching node
 | branching position
 | branching positions

\knowledge{notion}
 | Pick-priority nodes
 | pick-priority nodes
 | pick-priority node
 | pick-priority

\knowledge{notion}
 | coding nodes
 | $i$-coding nodes
 | $(i+1)$-coding nodes
 | $(i+1)$-coding
 | $i$-coding node
 | $\imax $-coding nodes
 | $\dmax $-coding nodes
 | $\imax $-coding node
 | $i$-coding
 | coding
 | $i_0$-coding
 | $\jmax $-coding
 | $r$-coding
 | $m$-coding
 | $i+1$-coding
 | $\imax $-coding
 | $j$-coding
 | $j$-coding node
 | $m$-coding node
 | $m'$-coding
 | $j$-coding nodes
 | $(j+1)$-coding nodes
 | $\imax +1$-coding
 | $\dmax $-coding
 | $(k+1)$-coding

\knowledge{notion}
 | memory update
 | update the memory

\knowledge{notion}
 | states
 | state

\knowledge{notion}
 | allows priority $3$
 | allow priority $3$

\knowledge{notion}
 | $C_{s-1}$-descendant
 | $C_{s-1}$-descendants

\knowledge{notion}
 | witnessing function

\knowledge{notion}
 | $i$-realizable
 | $\jmax $-realizable

\knowledge{notion}
 | type

\knowledge{notion}
 | $i$-consistent
 | $t$-consistent
 | $j$-consistent
 | $1$-consistent
 | $r$-consistent
 | $\sigma $-consistent
 | $\imax $-consistent
 | $\dmax $-consistent

\knowledge{notion}
 | has not been coded
 | not coded
 | was coded
 | is coded
 | has been coded

\knowledge{notion}
 | demands to code
 | demand to code

\knowledge{notion}
 | demand to witness
 | demands to witness

\knowledge{notion}
 | attainable

\knowledge{notion}
 | unattainable

\knowledge{notion}
 | recorded

\knowledge{notion}
 | primary
 | primary child

\knowledge{notion}
 | secondary
 | secondary child

\usepackage{macros}

\title{Automata for MSO over Infinite Trees with Quantification over Borel Sets of Branches}
\titlerunning{Automata for MSO over Infinite Trees with Quantification over Borel Sets\dots}

\author{Mikołaj Bojańczyk}{Institute of Informatics, University of Warsaw, Poland \and \url{https://www.mimuw.edu.pl/~bojan/}}{bojan@mimuw.edu.pl}{https://orcid.org/0000-0002-7758-1072}{Author supported by the Polish National Science Centre (NCN) grant ``Polynomial finite state computation'' (2022/46/A/ST6/00072).}

\author{Antonio Casares}{University of Kaiserslautern-Landau, Germany \and \url{https://antonio-casares.github.io/}}{antonio.casares@rptu.de}{https://orcid.org/0000-0002-6539-2020}{Partially supported by Deutsche Forschungsgemeinschaft (grant number 522843867) and European Research Council (grant number 101089343).
Part of this work was done while Casares was at the University of Warsaw, Poland, supported by the Polish National Science Centre (NCN) grant ``Polynomial finite state computation'' (2022/46/A/ST6/00072).}

\author{Sven Manthe}{University of Bonn, Germany}{sven.manthe@uni-muenster.de}{https://orcid.org/0009-0003-6108-6119}{Author supported by Hausdorff Center for Mathematics at the University of Bonn (DFG EXC 2047).}

\author{Paweł Parys}{Institute of Informatics, University of Warsaw, Poland \and \url{https://www.mimuw.edu.pl/~parys/}}{parys@mimuw.edu.pl}{https://orcid.org/0000-0001-7247-1408}{Author supported by the Polish National Science Centre (NCN) grant ``Frontiers of automatic analysis of concurrent systems'' (2021/41/B/ST6/00535).}

\authorrunning{M. Bojańczyk, A. Casares, S. Manthe, and P. Parys} 

\Copyright{Mikołaj Bojańczyk, Antonio Casares, Sven Manthe, and Paweł Parys} 



\category{} 

\relatedversion{} 

\nolinenumbers 

\begin{document}

\maketitle

\begin{abstract}
  Rabin's Tree Theorem says that the \mso theory of the infinite binary tree $2^*$ is decidable. Shelah showed that \mso logic becomes undecidable if this tree is extended to $2^{\le \omega}$, i.e.~by allowing quantification over sets of infinite branches. A longstanding open problem is whether the decidability can be recovered in $2^{\le \omega}$ by restricting  set quantification to Borel sets. We make some progress in this direction, by identifying a suitable automaton model, and showing that most of the automata-theoretic approach to Rabin's Theorem can be extended to the new framework. The only missing part is a conjecture about finite-memory determinacy in certain games. This paper states and explores the conjecture. We prove it in some restricted cases, and give lower bounds on the memory required in those games.
\end{abstract}

\maketitle

\section{Introduction}
One of the landmark results of logic in computer science is Rabin's Tree Theorem~\cite{treeautomata}, which states the decidability of the \mso theory of the infinite binary tree, the set $2^*$ equipped with the prefix relation. This is a powerful theory---for instance, it interprets the \mso theory of the rational numbers with the usual order~\cite[Thm.~2.1]{treeautomata}, which is thus decidable---and it is close to the border of undecidability.
On the other side of this border we find the Cantor set $2^\omega$, equipped with the lexicographic order, and also the order of the reals. Both of these structures have an undecidable \mso theory (as shown by Shelah~\cite[Theorem 7.3]{shelahord}, assuming the Continuum Hypothesis; this assumption was later removed in~\cite{monadicundecid}, i.e., undecidability can be proved in \textsc{zfc} only). The undecidability proofs use sets that are complicated from a topological point of view, and Shelah conjectured~\cite[Conjectures 7A and 7B]{shelahord} (see also \cite[Open Question, p. 503]{monadicsurvey}\cite[2.22]{shelahquestions}) that decidability can be recovered by restricting set quantification to range over Borel sets (we use the name \emph{Borel \mso} for the resulting logic):
\begin{enumerate} 
    \item[\textcolor{lipicsGray}{\sffamily\bfseries\upshape\mathversion{bold}""A"".}] The "Borel \mso" theory of $2^{\leq \omega}$ equipped with the prefix and lexicographic orders is decidable.
 \item[\textcolor{lipicsGray}{\sffamily\bfseries\upshape\mathversion{bold}""B"".}] The "Borel \mso" theory of $\mathbb R$ with the usual order is decidable.
\end{enumerate}
Conjecture B is the weaker one, since an isomorphic copy of the real numbers can be defined using "Borel \mso" in $2^{\le\omega}$.
In 2024, Manthe announced a proof of the weaker conjecture~\cite{Manthe24BorelMSO}. A proof of both conjectures was announced by Džamonja in~2025~\cite{Dzamonja25BorelMSO}, but we could not confirm its correctness.

This paper is devoted to the stronger Conjecture A. One can regard this conjecture as a common generalisation of Rabin's Tree Theorem and Conjecture B. Namely, the complete binary tree is recovered in $2^{\leq\omega}$ as the definable substructure consisting only of nodes in $2^*$.
Hence, any proof of Conjecture A would also need to subsume a proof of Rabin's Tree Theorem. In particular, this seems to rule out the proof of Manthe, which is based on linear orders and the compositionality of their \mso theories, and does not seem to be amenable to extensions that cover the tree structure. This paper proposes a programme to prove Conjecture A. Our main contributions can be summarised as follows:
\begin{enumerate}
    \item We introduce a certain kind of infinite games, and conjecture that these games can be won with finite memory.
    \item We prove that if the conjecture on finite memory is true, then Conjecture A is also true, that is, one can decide the "Borel \mso" theory of $2^{\leq \omega}$, and furthermore the logic has a corresponding automaton model.
    \item We present a preliminary study of the conjecture on finite memory, with partial positive results.
\end{enumerate}

Let us describe these contributions in some more detail.
Simply put, our programme is to copy the proof of Rabin's Tree Theorem, and extend it to account for Borel sets of branches. We begin by briefly recalling how Rabin's Tree Theorem is proved, using a modern presentation that emphasises games, as proposed by Gurevich and Harrington~\cite{GurevichHarrington82}. In the proof, one shows that every \mso formula can be transformed into an equivalent automaton, and the latter can be tested for emptiness. Proving this requires to show that automata have the same closure properties as the logic, namely Boolean combinations and projection. Union and intersection are easily achieved using a product construction, and projection is also easy because the automaton model in question is nondeterministic. The difficult closure property is complementation. Unlike for $\omega$\nobreakdash-words, determinisation is not an option. To prove complementation, one describes rejection by the automaton in terms of a game of infinite duration, and then proves that this game can be won using simple strategies of finite memory. The finite memory can then be embedded into the state space of a nondeterministic automaton that recognises the complementary language.

As we describe in this paper, most of the proof discussed in the previous paragraph can be adapted to the more general setting of $2^{\leq \omega}$. In this more general setting, an input tree has type $2^{\leq \omega} \to \Sigma$, that is, it assigns labels to both nodes and branches, and a run of the automaton has type $2^{\leq \omega} \to Q$. In both cases, we require the inverse image of every label to be Borel, since on the logical side we are targeting Borel \mso. The acceptance condition for the automaton is a variant of the Muller condition that takes into account the states assigned to the branches. Using this automaton model, we implement all parts of the proof of Rabin's Tree Theorem, with one exception, namely finite-memory determinacy of the underlying games. This determinacy result is the content of our conjecture, and much of the paper is devoted to its study.

\subparagraph*{The conjecture.} Let us now describe in more detail our conjecture on finite-memory strategies. Like most games studied in the context of automata, such as parity games or Muller games, our games are infinite duration games played by two players, called Adam and Eve, using directed graphs (called arenas), where each position is owned by one of the two players. This notion of game is standard, and will be recalled later in the paper. In this introduction, we focus on the distinguishing features of the games that arise within the context of Borel \mso, which are defined below. We call the games Muller-Borel, because the acceptance condition is a Boolean combination of \buchi and Borel conditions, and Boolean combinations of \buchi conditions are called Muller conditions.

\begin{definition}\label{def:muller-borel} \AP A ""Muller-Borel game"" is a game satisfying the following two conditions:
    \begin{enumerate}
        \item \textbf{Tree-like arena.} \label{item:tree-like-arena} The set of positions is $2^* \times Q$
        for a finite set $Q$. Edges of the game can only move to child nodes, that is, 
        if there is an edge from \((v,q)\) to \((w,p)\), then $w$ is a child of $v$. Not all edges that move to children need to be available.
        \item \textbf{Muller-Borel condition.} \label{item:muller-borel-condition} The winning condition is a finite Boolean combination of two kinds of conditions:
        \begin{description}
            \item[(\buchi)] \label{item:state-muller} $q \in Q$ is seen infinitely often;
            \item[(Borel)] \label{item:branch-borel} the branch belongs to a Borel set $B \subseteq 2^\omega$.
        \end{description}
        \end{enumerate}
\end{definition}

Using Martin's Borel Determinacy Theorem~\cite[p. 371]{Martin75}, one can show that the games are determined, that is, one of the players must have a winning strategy. However, our main interest is in the memory used by the winning strategy, which is beyond the scope of Martin's theorem. Our conjecture is that the games can be won with finite memory, as stated below. (The notion of finite memory is the standard one, and will be recalled later in the paper.)

\begin{conjecture}\label{conj:muller-borel}\label{conj:prioritized-borel}
    For every "Muller-Borel game", one of the players has a winning strategy that uses "finite memory". Furthermore,
    an upper bound on the memory size can be computed based on the size of~$Q$.
\end{conjecture}

Using the same framework as for Rabin's Tree Theorem, we prove in \cref{thm:consequences-of-conjecture} that the above conjecture implies decidability of "Borel \mso" on $2^{\leq \omega}$, that is, Shelah's Conjecture A. The rest of this introduction is devoted to a preliminary discussion of the conjecture, with an extended discussion contained in the main body of the paper.

As indicated, the main examples of Muller-Borel games appear in the analysis of tree automata. One can recognize these games in the tree-like requirement (\cref{item:tree-like-arena} of \cref{def:muller-borel}), with $Q$ being the set of states.
For many kinds of games, such as Muller games or parity games, the tree-like shape of the arena is irrelevant in the analysis of the game.
Hence, those games are commonly defined over arbitrary graphs~\cite{Thomas95,BookGamesOnGraphs2025}. In our case, however, the tree-like shape is essential, since the Borel conditions in the game are defined in terms of the tree structure. Let us now explain how certain special cases of a Muller-Borel game correspond to important results on determinacy of games.
\begin{itemize}
    \item Suppose that we only allow \buchi conditions in the game. In this case, we have a game with a Muller condition. Muller games are known to have finite-memory strategies, which was first proved by \buchi and Landweber for finite arenas~\cite[Theorem 1']{BL69Strategies}, and then by Gurevich and Harrington for general arenas~\cite[Theorem 1]{GurevichHarrington82}. Therefore, if we only allow \buchi conditions, then the conjecture is true.
    \item Suppose that there is only one state in the game. In this case, each game position can be reached in only one way (since the positions form a tree), and therefore all strategies are memoryless. Thus, from Martin's Borel Determinacy Theorem we can conclude that the conjecture holds in this special case, even without any memory. The full power of Martin's theorem is needed for this special case.
\end{itemize}

Later on in the paper, we prove the conjecture for other special cases of the game, including the special case where only the Borel condition is allowed (extending the situation from the second item above, in which the \buchi condition trivializes).

Our conjecture considers finite-memory strategies. A natural question would be about memoryless strategies. For example, in automata theory, finite-memory determinacy of Muller games is commonly proved by first reducing to a parity condition, and then proving that parity games can be won using memoryless strategies. The only need for memory arises from the reduction from Muller to parity, which is typically accomplished using a variant of McNaughton's ``latest appearance record'' construction. For our conjecture, we are unable to present such a reduction, that is, we are not aware of a variant of our conjecture which would talk about memoryless strategies. Using priorities does not seem to help: memory is needed even if we convert the game into a parity-style format, in which the \buchi part of the acceptance condition is replaced by priorities on edges. In fact, even in the presence of only three priorities, the amount of needed memory grows unboundedly as a function of the number of states (see \cref{prop:unbounded-memory}). On the positive side, we show that for two priorities, only two states of memory are needed, regardless of the number of states in $Q$ (see \cref{prop:two-priorities}).
Other lower and upper bounds are contained in \cref{prop:one-priority,prop:nested,prop:k-copying-game}.

The conjecture seems rather strong. In particular, as we discuss in \cref{sec:f-sigma}, the conjecture implies that if we restrict the semantics of "Borel \mso" to use only Boolean combinations of $\Fsigma$ sets (i.e. countable unions of closed sets), then the same formulae are true.
In other words, "Borel \mso" can essentially only talk about Boolean combinations of $\Fsigma$ sets, and it cannot access more complicated Borel sets. Since one could expect "Borel \mso" to be more expressive, this could be seen as evidence against the conjecture. However, the recent proof of Manthe is consistent with our conjecture, since it implies that in the weaker setting of $2^\omega$ one can also restrict quantification to Boolean combinations of $\Fsigma$ sets, and the same formulae are true.

\subparagraph*{Lean formalization.}
The vast majority of the results proved in this paper have been formalized in Lean. The formalization is available as ancillary files for the arXiv version (v1) of this article. Further details are provided in the \texttt{README.md} file included with the formalization.

\subparagraph*{Changes from the conference version.}
This version contains several corrections and additions. We have added \cref{prop:single-state}, showing that in one-state prioritized Borel games with \(k\) priorities the winner can use at most \(k\) memory states, matching the lower bound now stated as \cref{prop:k-copying-game}. We have corrected \cref{prop:unbounded-memory} (Proposition 28 in the conference version): the game used in the previous proof can in fact be won with two memory states and therefore does not establish the claimed lower bound. We have also corrected \cref{remark:topological-simplification}, which contained an incorrect topological claim, and the proof of \cref{prop:nested}. In addition, we have refined the topology on \(2^{\leq\omega}\) and the formalism for game arenas, and have expanded and clarified several arguments throughout the paper. Finally, we have added a Lean formalization of the results as ancillary material.

\subparagraph*{Use of AI.}
No AI tools were used in the research leading to, or in the writing of, the first version of this paper. In revising the paper, Claude Fable 5 and Opus 5 helped identify the error in the proof of \cref{prop:unbounded-memory} while analyzing games using a SAT solver. ChatGPT was used to assist with the formulation of some passages. The Lean formalization was developed using the Aristotle tool, and this process also helped uncover several minor errors in the paper.

\section{Borel MSO}

In this section, we describe the logics in question. We begin with the usual notion of monadic second-order logic (\mso). In this logic, there are two kinds of variables: lowercase variables $x,y,z$ for elements and uppercase variables $X,Y,Z$ for sets. Formulas are built using the following constructors:
\begin{align*}
    \hskip1.5cm
\myunderbrace{\exists x \ \forall x}{quantification \\ over elements}
\qquad \quad
\myunderbrace{\exists X \ \forall X}{quantification \\ over sets}
\qquad \quad
\myunderbrace{\lor\ \land \ \neg}{Boolean \\ combinations}
\qquad
\myunderbrace{R(x_1,\ldots,x_n)}{relations from\\ the vocabulary}
\qquad
\myunderbrace{x\in X}{membership\\ relation}.
\end{align*}
The semantics of the logic are defined in the usual way (see e.g.~\cite{Thomas97} for a more in-depth discussion).
When we talk about the \mso theory of a structure, we mean the set of \mso sentences that are true in this structure.
Rabin's Tree Theorem~\cite{treeautomata} states the decidability of the \mso theory of the infinite binary tree, which is viewed as a structure where the underlying set is $2^*$, and there are two binary relations: the prefix order, and the lexicographic order.

\AP The logic that is studied in this paper is ""Borel \mso"". This logic has the same syntax as \mso, except that the semantics are changed so that set quantification is restricted to range only over Borel sets. Recall that the family of Borel sets in a topological space is the least family of subsets that contains the open sets, and which is closed under complementation as well as under taking countable unions and intersections. Therefore, in order to speak of Borel \mso, the underlying set in a structure must be equipped with a topology. The case which interests us is where the underlying set is $2^{\leq \omega}$, that is, strings of bits that have length at most $\omega$.
\AP The finite strings are called ""nodes"" and the infinite ones are called ""branches"".
\AP The topology is the one which arises taking as base of open sets the sets of the form
\[S_w = \{w\}\quad\text{and}\quad U_w = \{x\in 2^\omega \mid w \text{ is a prefix of } x\} , \quad \text{where } w\in 2^*. \]
That is, the ""open sets"" of this topology are the (arbitrary) unions of $S_w$ and $U_w$, for $w\in 2^*$.

When we speak of $2^{\leq \omega}$ as a structure, we assume that it is equipped with the prefix and lexicographic orders, just as in Rabin's Tree Theorem. The prefix relation gives us the tree structure on nodes and allows us to check whether a node belongs to a branch, while the lexicographic order distinguishes the left subtree from the right one.

This paper is devoted to "Conjecture A" from the introduction, which says that the "Borel \mso" theory of $2^{\leq \omega}$ is decidable, meaning that there is an algorithm that inputs a sentence of the logic, and answers whether the sentence is true in the structure. Note that the Borel restriction is relevant only for branches. This is because every set of nodes is Borel, and thus a subset of $2^{\leq \omega}$ is Borel if and only if its restriction to branches is a Borel set.

\begin{example}[Limit points] \label{ex:nodes-vs-branches} In this example, we give an $\mso$ sentence distinguishing the structures $2^*$ and $2^{\leq \omega}$.
It suffices to consider the sentence:
\[ \forall X. \ X \text{ is a chain for the prefix order} \ \rightarrow \ \exists y.\ \forall x\in X.\ x \text{ is a prefix of } y.\]
This does not hold in $2^*$, taking $X$ to be all nodes on an arbitrary branch.
However, it holds in $2^{\leq \omega}$, as the branch itself is an element of the structure, so it provides an upper bound to~$X$.
(Note the similarities with the sentence  ``every bounded sequence admits a limit point'' distinguishing $\Q$ and $\R$.)
\end{example}

\begin{example}[Determinacy] \label{ex:determinacy} In this example, we illustrate the difference between \mso and Borel \mso over $2^{\leq \omega}$.
    A Gale-Stewart game is a game where two players, call them Adam and Eve, pick bits from $2=\set{0,1}$ in alternation. After $\omega$ rounds, an infinite play in $2^\omega$ arises. The winner of the game is given by a winning condition
     $W \subseteq 2^\omega$, which tells us which plays are winning for Eve. A strategy for player Eve can be described as a set of nodes $X$, which indicates the chosen successors used by the strategy. Similarly, a strategy for Adam can be described as a set of nodes $Y$. Therefore, one can express determinacy of such games using a formula:
     \begin{align*}
     \forall W \subseteq 2^\omega. \
     \lor
     \begin{cases}
     \exists X \subseteq 2^*. \ \forall Y \subseteq 2^*. \ \varphi(W,X,Y) \\
      \exists Y \subseteq 2^*. \ \forall X \subseteq 2^*. \ \neg \varphi(W,X,Y),
     \end{cases}
     \end{align*}
     where $\varphi(W,X,Y)$ says that the unique play consistent with strategies $X$ and $Y$ belongs to the set~$W$. If we think of this formula as a formula of Borel \mso, then it is true, since only Borel winning conditions are considered, and Martin's Borel Determinacy Theorem says that such games are determined. If we think of this formula as a formula of (unrestricted) \mso, then it is false, since there exist games with (non-Borel) winning conditions that are not determined, assuming the Axiom of Choice.
\end{example}


\section{Nondeterministic automata over Borel trees}

\newcommand{\boreltrees}[1]{\mathsf T #1}

The idea behind the automata method is to find an automaton model that recognises exactly the same languages as those that can be defined in Borel \mso. These languages consist of labelled trees. The intuitive idea is that the labels describe a valuation of free variables for a formula that has free variables.
Let us begin by describing the appropriate notion of trees.

\begin{definition}[Borel tree]\label{def:borel-tree}
    \AP A ""Borel tree"" over a finite alphabet $\Sigma$ is defined to be a function
\begin{align*}
t \colon 2^{\leq \omega}\to \Sigma
\end{align*}
such that for every letter $a \in \Sigma$, the inverse image is a Borel set. We denote the set of "Borel trees" over the alphabet $\Sigma$ by $\boreltrees{\Sigma}$.
\end{definition}
 Since the topology is only important on the infinite branches, the Borel condition is only important for those, and the labels of the nodes are unconstrained.

 As mentioned earlier in this section, we consider languages of "Borel trees" that can be defined in logic, or recognised by automata. Let us first explain how a language of "Borel trees" can be defined in logic.
 We say that a language $L \subseteq \boreltrees\Sigma$ for $\Sigma=\set{a_1,\ldots,a_n}$ is definable in "Borel \mso" if there is a formula
 \begin{align*}
 \varphi(X_1,\ldots,X_n)
 \end{align*}
 of "Borel \mso" such that a tree $t \in \boreltrees \Sigma$ belongs to $L$ if and only if the formula is true under the valuation that maps each variable $X_i$ to the set of nodes and branches that are labelled by the letter $a_i$. Our goal in this section is to define an automaton model which recognises exactly the languages definable in Borel \mso. (This goal is achieved conditionally on a certain conjecture about complementation of automata.)

\subparagraph*{Automata over "Borel trees".} We now describe the automaton model that is used to recognise languages of "Borel trees".
 This model is a variant of nondeterministic Muller tree automata.
\begin{definition}[Nondeterministic automaton]\label{def:nondet-borel-automaton}
    \AP A ""nondeterministic Muller automaton on Borel trees"" is given by:
    \begin{itemize}
        \item a finite set of states $Q$;
        \item a finite alphabet $\Sigma$;
        \item a subset $I \subseteq Q$ of initial states;
        \item two transition relations:
        \begin{align*}
        \Delta_* \subseteq Q \times \Sigma \times Q \times Q,
        \qquad
        \Delta_\omega \subseteq \powerset Q \times \Sigma \times Q.
        \end{align*}
    \end{itemize}
\end{definition}

An input "Borel tree"
$
t \in \boreltrees \Sigma
$
is accepted by the automaton if there exists an accepting run, which is defined to be a Borel tree
$
\rho \in \boreltrees Q
$
such that the following three conditions hold:
\begin{enumerate}[(1)]
    \item the root is labelled by an initial state;
    \item for every node $x \in 2^*$, the transition relation $\Delta_*$ contains the tuple consisting of: the state in $x$, the label of $x$ in $t$, the state in the left child of $x$, the state in the right child of $x$;
    \item for every branch $b \in 2^\omega$, the transition relation $\Delta_\omega$ contains the tuple consisting of: the set of states that occur infinitely often along $b$, the label of $b$ in $t$, the state in $b$.
\end{enumerate}

\begin{lemma}\label{lem:borel-emptiness}
    Emptiness is decidable for "nondeterministic Muller automata on Borel trees".
\end{lemma}

\begin{proof} We reduce the problem to emptiness for nondeterministic Muller automata on the usual kind of infinite trees, which do not have labels or states on branches. For the latter kind of automata, emptiness is decidable.

    Consider a "nondeterministic Muller automaton on Borel trees".
    The "Muller automaton on Borel trees" is nonempty if and only if there is a labelling $\rho : 2^* \to Q$ which satisfies the three conditions (1), (2), (3) in the definition of an accepting run, except that condition (3) is changed so that for each branch $b \in 2^\omega$, the set of states that occur infinitely often along $b$ belongs to
        \begin{align*}
    \mathcal F = \set{ R \subseteq Q\mid \text{$(R,a,q) \in \Delta_\omega$ for some $a \in \Sigma$ and $q \in Q$}}.
    \end{align*}
    We define a Muller tree automaton on infinite trees in the usual sense, with the same set of states and transitions on finite nodes, and with \(\Ff\) as its acceptance condition.
    
    This new automaton is nonempty if and only if the original "Borel tree" automaton is nonempty. Indeed, if the original automaton accepts some Borel tree, then by restricting the input tree and its accepting run to finite nodes we obtain an accepting run of the usual Muller automaton. Conversely, suppose that the usual Muller automaton accepts some infinite tree. For every \(R\in \Ff\), fix an input letter \(a\in\Sigma\) and a state \(q\in Q\) such that \((R,a,q)\in\Delta_\omega\). On each branch whose set of states occurring infinitely often is \(R\), assign the chosen letter \(a\) to the input tree and the chosen state \(q\) to the run. The set of branches labelled by a particular input letter or by a particular state is Borel, since it is a finite union of sets of branches on which exactly the states from some \(R\subseteq Q\) occur infinitely often; each such set is Borel, being defined by finitely many Büchi and co-Büchi conditions. Thus, the input tree and the run extend to Borel trees satisfying the acceptance condition of the original automaton.
\end{proof}

We note that atomic $\mso$ formulas (prefix and lexicographic orders and membership relation) can be encoded into Muller automata on Borel trees just as they can be translated into classical Muller automata. The only subtlety in such encoding, which already arises in the classical setting of $2^*$, is the handling of free variables: a tree with an evaluation of its free variables can be encoded using an extended alphabet.

We now establish some straightforward closure properties of our automata, namely union, intersection, and projection. The difficult part is complementation. We are unable to prove complementation, but the rest of this paper is devoted to a discussion of this problem, and some possible approaches towards its solution.

Let us begin with union and intersection.
\begin{lemma}\label{lem:union-and-intersection}
    Languages recognised by "nondeterministic Muller automata on Borel trees" are closed under union and intersection.
\end{lemma}
\begin{proof}
    Standard product construction. This construction works because the Borel condition for trees is closed under pairing. This means that if we take two trees over alphabets $\Sigma_1$ and $\Sigma_2$ and pair them into a single tree over the alphabet $\Sigma_1 \times \Sigma_2$, then the new tree is a "Borel tree" if and only if both of the original trees were "Borel trees".
\end{proof}

We now turn to projection, which corresponds to existential set quantification in logic.
Since our automaton model is nondeterministic, it is also closed under taking projections, understood as follows. Consider a function
\begin{align*}
f\colon \Sigma \to \Gamma
\end{align*}
from one finite alphabet to another. This function can be extended to a function on Borel trees
\begin{align*}
\boreltrees f \colon \boreltrees \Sigma \to \boreltrees \Gamma
\end{align*}
in the obvious way, by applying $f$ to the label of every node and branch. We can apply this function to some language $L \subseteq \boreltrees \Sigma$, obtaining a new language $f(L) \subseteq \boreltrees \Gamma$, which is called the \emph{projection} of $L$ along~$f$.
\begin{lemma}\label{lem:nondeterministic-projection}
    Languages recognised by "nondeterministic Muller automata on Borel trees" are closed under taking projections.
\end{lemma}
\begin{proof}
    As usual, given a projection \(f:\Sigma\to\Gamma\), we obtain an automaton over \(\Gamma\) by allowing, for every input letter in \(\Gamma\), any transition of the original automaton whose input letter in \(\Sigma\) projects to it.

    If the original automaton accepts a Borel tree, then the projected automaton accepts its projection. Indeed, the same run can be used, and the projected tree is Borel because applying \(f\) to a Borel labelling preserves Borelness.

    For the converse direction, the only point that requires attention is Borelness. Suppose that the projected automaton has an accepting Borel run. At every finite node, choose arbitrarily an input letter in \(\Sigma\) witnessing the chosen transition; there is no Borel restriction on labels of finite nodes. On branches, fix in advance, for every possible set of states occurring infinitely often, input letter in \(\Gamma\), and branch state, one input letter in \(\Sigma\) witnessing the corresponding infinitary transition. The set of branches assigned a particular input letter is Borel. Indeed, it is a finite union of intersections of three kinds of Borel sets: branches on which exactly a given set of states occurs infinitely often, branches carrying a given input letter in \(\Gamma\), and branches carrying a given state in the run. The first kind is Borel because it is defined by finitely many Büchi and co-Büchi conditions, while the other two are Borel because the input tree and the run are Borel. We thus obtain a Borel tree over \(\Sigma\) which is accepted by the original automaton and projects to the given tree.
\end{proof}

As in the proofs of \buchi or Rabin, the essential step is complementation. We do not know how to prove this, and leave it as a conjecture. In the next section, we present a conjecture on finite-memory strategies in games, which implies that languages recognised by "nondeterministic Muller automata on Borel trees" are effectively closed under complementation, that is, for every automaton one can compute an automaton for its complement language. This in turn implies our main conjecture about decidability of "Borel \mso" (see \cref{thm:consequences-of-conjecture}).

\section{Complementation of automata and games}
When proving closure under complementation for nondeterministic automata on infinite trees, games and their determinacy play a central role. We follow that same path. Let us begin by recalling the kind of games that are used in our proofs.

\subsection{Games} The games that we consider have infinite duration, and are played by two players Adam and Eve on a directed graph. We now recall the definition of such games, and the basic concepts such as strategies.

Throughout the paper, we use directed graphs with a finite set of edge labels. Formally, such a graph consists of a set \(V\) of vertices, a finite set \(\Lambda\) of edge labels, and a relation \(E\subseteq V\times\Lambda\times V\). We write \(v\xrightarrow{\lambda}w\) if \((v,\lambda,w)\in E\). Paths, both finite and infinite, are defined in the usual way, including the edge labels. Whenever edge labels are not explicitly specified, we use a singleton set of edge labels.

\begin{definition}[Game]
    \AP A ""game"" is given by:
    \begin{enumerate}
        \item a directed graph $(V,\Lambda,E)$, called the ""arena"";
        \item an ownership function of type $V \to \set{\Adam,\Eve}$;
        \item an initial position $v_0 \in V$;
        \item a winning condition $W$, which assigns a winner in $\set{\Adam,\Eve}$ to every infinite path starting in~$v_0$.
    \end{enumerate}
\end{definition}

In the context of a game, vertices of the underlying graph are called ""positions"".
We assume that in the arena, every position has at least one outgoing edge.

We now briefly recall the usual notions of strategies and winning strategies for games. \AP A ""strategy"" for player $i$ is a function that inputs a finite path in the "arena" from the initial position to some position owned by $i$, and outputs an edge that extends this path.
\AP In the following, we refer to such finite paths as ""histories"".
Once both players fix their strategies, a unique infinite path arises in the game, which we call a ""play"", and the winning condition tells us which of the two players wins.
Globally, a player wins if they have a winning strategy, that is, a strategy that wins against every strategy of the opponent.
\AP A game is ""determined"" if one of the players has a winning strategy. Under the Axiom of Choice, not every game is determined, but Martin's Borel Determinacy Theorem says that the game is determined if the winning condition is a Borel subset of all paths, according to the usual topology (which, in particular, is the case for games arising in this paper).

\AP We are particularly interested in ""finite-memory strategies"", which are defined as follows. Essentially, these are strategies that are produced by finite state transducers reading the edges in the game.
\AP More formally, a strategy for Eve (for Adam, the concept is defined symmetrically) with memory $M$ consists of some initial memory state $m_0$ and two functions
\begin{align*}
\myunderbrace{\delta\colon M \times E \to M}{memory update}
\quad \text{and} \quad \myunderbrace{\sigma\colon M\times \myoverbrace{V_{\Eve}}{positions owned by Eve} \to E}{Eve's choice},
\end{align*}
where $\sigma(m,v)$ is required to have source $v$.
The first function is used to ""update the memory"" along a "play" (as in a deterministic automaton) and the second function shows how Eve chooses edges in the game from her positions.
\AP A ""finite-memory strategy"" is one where the memory set $M$ is finite.
\AP A ""memoryless strategy"" is the special case of a strategy with finite memory in which the set $M$ has one element only. A game has \emph{finite-memory determinacy} if one of the two players can win using a strategy with finite memory.
It was shown by Gurevich and Harrington~\cite[Theorem 1]{GurevichHarrington82} that games with a Muller winning condition (which topologically speaking is a finite Boolean combination of $\Fsigma$ sets) have finite-memory determinacy.

The two determinacy results discussed above are incomparable. Indeed, Martin's Borel Determinacy Theorem has a weaker assumption (the winning condition is Borel), but a weaker conclusion (the winning strategy might use infinite memory). Both results play an important role in our setting, since our games are a hybrid of Borel and Muller games.

\subsection{Complementation}
\label{sec:alternating-automata}
To complement nondeterministic automata, we use a game based approach, following Gurevich and Harrington~\cite{GurevichHarrington82}.
In this approach, for a nondeterministic automaton $\Aa$ and an input tree $t$, we define an \emph{acceptance game} as follows. In the game, Eve tries to show that the tree is accepted, by constructing an accepting run, and Adam tries to dispute Eve's claim, by showing that her run breaks the infinitary part $\Delta_\omega$ of the transition relation. We assume without loss of generality that the automaton has at least one transition from every state for every input letter, by adding a rejecting sink state if necessary. The positions of the game are then pairs in
\begin{align*}
\hspace{0.9cm}\myunderbrace{(Q + Q^2)}{disjoint union of \\ states and pairs of states}{} \times 2^*.
\end{align*}
The initial position consists of the initial state and the root of the tree. (One can assume without loss of generality that the automaton has one initial state.) In a position \((q,v)\) with \(q\in Q\), Eve chooses a pair of states \((q_0,q_1)\in Q^2\) such that \((q,t(v),q_0,q_1)\in\Delta_*\). In the resulting position, Adam picks a child direction $i \in 2$ and the game proceeds to the position $(q_i, v \cdot i)$. The winning condition in the game is defined as follows. Suppose that $R \subseteq Q$ is the set of automaton states that are visited infinitely often in the play, and $b \in 2^\omega$ is the branch that is traced by this play in the underlying tree. Then Eve wins the play if and only if the branch $b$ belongs to the set
\begin{align*}
\set{ b \in 2^\omega \mid \text{there is a state $q$ such that $(R,t(b),q) \in \Delta_\omega$}}.
\end{align*}
For every fixed \(R\subseteq Q\), the above set is Borel, since it is a finite union of sets of branches carrying a particular input letter. It follows that the acceptance game is almost a special case of a "Muller-Borel game", with the only minor issue being that the "arena" is not strictly speaking tree shaped, since the play can stay in the same node for two consecutive positions. This minor issue will be resolved shortly (see \cref{claim:minor}).

\begin{lemma}
    A tree is accepted by a nondeterministic automaton if and only if Eve has a winning strategy for the corresponding acceptance game.
\end{lemma}

\begin{proof}
    From an accepting run, we easily construct a winning strategy for Eve which follows this run, and vice versa. The only issue that requires some attention is that when we convert a winning strategy into a run, we need to ensure that the states in the branches are defined in a Borel way. To this end, consider a winning strategy $\sigma$ for Eve. For every \(R\subseteq Q\) and \(a\in\Sigma\) for which there exists a state \(q\in Q\) such that \((R,a,q)\in\Delta_\omega\), fix one such state \(q\). For every branch \(b\in2^\omega\), consider the unique play that is consistent with \(\sigma\) and traces \(b\); let \(R\) be the set of states occurring infinitely often in this play. We assign to \(b\) the state fixed for the pair \((R,t(b))\); such a state exists because \(\sigma\) is winning. Moreover, for every state \(q\), the set of branches assigned state \(q\) is Borel, since it is a finite union of intersections of sets of branches on which exactly the states from some \(R\subseteq Q\) occur infinitely often with sets of branches carrying a particular input letter. The former sets are Borel, being defined by finitely many Büchi and co-Büchi conditions, while the latter are Borel because the input tree is Borel. Thus, the resulting run is a Borel tree and is accepting.
\end{proof}

The acceptance game is designed so that a tree is rejected if and only if player Adam has a winning strategy in the acceptance game. As we show in the following lemma, this can be expressed by a nondeterministic automaton, provided that the winning strategy has finite memory, as ensured by \cref{conj:muller-borel}.

\begin{lemma}\label{lem:complementation} Assume \cref{conj:muller-borel}. Then languages recognised by "nondeterministic Muller automata on Borel trees" are closed under complementation, and the complement automaton can be effectively computed.
\end{lemma}
\begin{proof}
    Fix a nondeterministic Muller automaton on "Borel trees" $\Aa$. We want to compute an automaton, of the same kind, for its complement language. This automaton will guess a finite-memory winning strategy for Adam in the corresponding acceptance game. Such a strategy exists for trees in the complement language by \cref{conj:muller-borel}, as shown in the following claim.

    \begin{claim}\label{claim:minor}
        Assume \cref{conj:muller-borel}. Based on the state space $Q$ of $\Aa$ one can compute a memory bound $m$ such that for every input tree, the corresponding acceptance game can be won using a strategy with at most $m$ memory states.
    \end{claim}
    \begin{claimproof}
        This is essentially the content of the conjecture, with the minor issue that the games from the conjecture require that one always moves to a child, while in the acceptance game Eve's move from a state to a pair of states stays in the same tree node.
        This can be easily resolved by modifying the game so that this move always goes to the left child, and suitably updating the Borel part to ignore these artificial directions. The resulting "Muller-Borel game" has \(|Q|+|Q|^2\) states, so the memory bound provided by \cref{conj:muller-borel} can be computed based only on \(Q\).
    \end{claimproof}

    Let $M = \set{1,\ldots,m}$ be the memory that is needed according to the above claim. By the claim it follows that an input tree $t$ is rejected by the automaton $\Aa$ if and only if there exists a winning strategy of player Adam in the acceptance game that uses memory $M$. Such a strategy can be viewed as a function that assigns two pieces of information to each node $v \in 2^*$ in the input tree, namely the functions
    \begin{align*}
    \myunderbrace{\delta_v\colon M \times (Q^3 + Q^2\times 2) \to M}{memory update} \quad\! \text{and} \quad\!
    \myunderbrace{\sigma_v\colon M\times Q^2 \to 2}{Adam's choice}.
    \end{align*}
    In the memory update \(\delta_v\), the two summands correspond to the two kinds of edges in the acceptance game: an Eve edge is determined, at a fixed tree node, by the source state and the pair of states chosen by Eve, while an Adam edge is determined by this pair together with the chosen direction. The function \(\sigma_v\) describes Adam's choice of direction from a position represented by a pair of states. A nondeterministic automaton can guess such a strategy, since a function as above can be thought of as a tree over alphabet
    \begin{align*}
        \Gamma \eqdef (M \times (Q^3 + Q^2\times 2) \to M) \times (M\times Q^2 \to 2) + \set{\omega},
    \end{align*}
    where the label $\omega$ is used for branches. Below, we show that once an input tree and a strategy for Adam are given in this representation, then a nondeterministic automaton can check if the strategy is winning for Adam. In other words, the following language, which we call the \emph{strategy language}, is recognised by a nondeterministic automaton (in fact, a deterministic one would be enough):
    \begin{align*}
    \setbuild{ t \in \boreltrees{(\Sigma \times \Gamma)}}{the $\Gamma$ part represents a winning strategy for \\ Adam in the acceptance game for the $\Sigma$ part}.
    \end{align*}
    This completes the complementation proof, since by projecting the strategy language to $\boreltrees{\Sigma}$, we get the complement of the language accepted by $\Aa$.

    To define an automaton for the strategy language, we observe that it only checks a certain $\omega$-regular property on all branches. This is because a tree belongs to the strategy language if and only if: for every branch $b \in 2^\omega$, (*) if we take any play of the acceptance game that is consistent with Adam's strategy and which stays on the branch $b$, then this play is winning for Adam. To determine whether a play is winning for Adam, we need access to the label of the branch $b$ in the input tree, since this label is used in the winning condition. Once we know this label, condition (*) becomes an $\omega$-regular property of the branch, since it can be defined in \mso over $\omega$-words. Therefore, to show that the strategy language is recognised by an automaton, we can use the following claim, where the alphabet $\Sigma$ is the alphabet $\Sigma \times \Gamma$ of the strategy language.
    \begin{claim} Let $\Sigma$ be a finite alphabet, and suppose that for each letter $a \in \Sigma$ we have an associated $\omega$-regular language $L_a$ over the alphabet $\Sigma \times 2$. Then we can construct a nondeterministic (even deterministic) Muller automaton over "Borel trees" that accepts a tree $t \in \boreltrees \Sigma$ if and only if for every branch $b \in 2^\omega$ with label $a$, the language $L_a$ contains the $\omega$-word given by the sequence
        \begin{align*}
          (\text{label of $t$ in $i$-th node on $b$}, \text{direction in $i$-th step of $b$})_{i=0}^\omega.
        \end{align*}
    \end{claim}
    \begin{claimproof}
        For each language $L_a$ take a deterministic Muller automaton on $\omega$-words that recognises it, and run these automata in parallel (using a product construction) along each branch. The infinitary part of the transition relation of the resulting tree automaton, having access to the label $t(b)$ of the branch and to the states of the automata that occur along this branch, checks whether the automaton for $L_{t(b)}$ accepts.
    \end{claimproof}
This concludes the proof of \cref{lem:complementation}.    
\end{proof}

Let us now put together all results proved above to show that \cref{conj:muller-borel} implies decidability of the "Borel \mso" theory of $2^{\leq \omega}$, as stated in Shelah's "Conjecture A".

\begin{theorem}\label{thm:consequences-of-conjecture}
   Assume \cref{conj:muller-borel}. Then "nondeterministic Muller automata on Borel trees" define the same languages as Borel \mso, and translations in both directions are effective. Furthermore, the "Borel \mso" theory of $2^{\leq \omega}$ is decidable.
\end{theorem}
\begin{proof}
  To go from an automaton to a formula of Borel \mso, one simply uses the logic to express the existence of an accepting run, in the usual way. To go in the opposite direction, we realize particular constructors of logical formulae using the closure properties that have been established above: intersections and unions (\cref{lem:union-and-intersection}), projections (\cref{lem:nondeterministic-projection}) and complementation (\cref{lem:complementation}).
  The conjecture is used only for complementation. Finally, to decide the "Borel \mso" theory, we convert a sentence of the logic into an equivalent automaton (over a single-letter alphabet), and then use \cref{lem:borel-emptiness} to check whether it is nonempty.
\end{proof}

\subsection{Boolean combinations of \texorpdfstring{$\mathbf{\Fsigma}$}{F-sigma} sets}
\label{sec:f-sigma}

\newcommand{\fsigmatree}{\mathsf S}

In this section, we discuss a consequence of our main conjecture, which is that "Borel \mso" is equivalent to a logic where quantification is much more restricted, namely to Boolean combinations of $\Fsigma$ sets.

\AP Recall that $\intro*\Fsigma$ sets are countable unions of closed sets; another notation for the same family of sets is level $\bsigma{2}$ of the Borel hierarchy.
In the topological space $2^{\omega}$, these are exactly the sets that can be described using a co-\buchi condition on the visited nodes, that is, as
\begin{align*}
X = \set{ b \in 2^\omega\mid\text{$b$ visits nodes from $F$ finitely often}}
\end{align*}
for some set of nodes $F \subseteq 2^*$; see~\cite[Corollary~3.2]{Landweber69OmegaAut}.
If we take a Boolean combination of such sets, we arrive at a Muller condition, that is, we have a finite family of sets $F \subseteq 2^*$, and whether  a branch belongs to $X$ depends on which sets $F$ are visited infinitely often.

One could consider a variant of  Borel \mso, which has the same syntax, but for which the semantics are changed so that set quantification is restricted to Boolean combinations of $\Fsigma$ sets.
In his proof of "Conjecture B", Manthe showed that over the real line, this new logic gives rise to the same theory, that is, restricting  quantification to  Boolean combinations of $\Fsigma$ sets does not change the truth value of sentences~\cite[Theorem~1.1]{Manthe24BorelMSO}.
The following theorem shows that, assuming our conjecture, the same result would hold for the structure $2^{\leq \omega}$.
\begin{theorem}\label{thm:f-sigma}
    Assume \cref{conj:muller-borel}.
    "Borel \mso" has the same true sentences as the restriction of \mso where set quantifiers range over Boolean combinations of $\Fsigma$ sets.
\end{theorem}
\begin{proof}
    Let us define $\fsigmatree \Sigma \subseteq \boreltrees{ \Sigma}$ to be the set of "Borel trees" where the labelling is restricted to Boolean combinations of $\Fsigma$ sets, that is, the inverse image of each label $a\in \Sigma$ is a Boolean combination of $\Fsigma$ sets. We prove that for every formula $\varphi$, which describes a property of trees labelled by an alphabet $\Sigma$, we have
    \begin{align*}\label{eq:two-semantics}
    \hspace{0.7cm}\myunderbrace{t \models_{\scriptscriptstyle\mathsf T} \varphi}{Borel semantics} \qquad \iff \qquad
    \myunderbrace{t \models_{\scriptscriptstyle\mathsf S} \varphi}{Bool($\Fsigma$) semantics}
    \qquad \text{for every tree }t \in \fsigmatree \Sigma.\tag{$\star$}
    \end{align*}
    The proof is a straightforward induction, and the only interesting case is that of a quantifier, say an existential one. To prove this step, we convert the quantified formula into an automaton using \cref{thm:consequences-of-conjecture}, and then we apply the following claim to show that the quantified set can always be chosen so that it is a Boolean combination of $\Fsigma$ sets.

    \begin{claim}
        Let $\Aa$ be  a nondeterministic Muller automaton over "Borel trees" with input alphabet $\Gamma$ and state space \(Q\), and let $f\colon \Gamma \to \Sigma$ be some projection function. For every tree $t \in \fsigmatree\Sigma$, the following conditions are equivalent:
        \begin{enumerate}
            \item\label{item:fsigma} $\Aa$ accepts some $s \in \fsigmatree \Gamma$ that projects to $t$;
            \item\label{item:borel} $\Aa$ accepts some $s \in \boreltrees \Gamma$ that projects to $t$.
        \end{enumerate}
    \end{claim}
    \begin{claimproof}
        The implication from \cref{item:fsigma} to \cref{item:borel} is immediate. For the converse, suppose that \(\Aa\) accepts some \(s\in\boreltrees\Gamma\) that projects to \(t\), and let \(\rho\) be an accepting run on \(s\). For every \(R\subseteq Q\) and \(a\in\Sigma\) for which there exist \(c\in\Gamma\) and \(q\in Q\) such that \(f(c)=a\) and \((R,c,q)\in\Delta_\omega\), fix one such pair \((c,q)\). We keep the labels of \(s\) and \(\rho\) on finite nodes unchanged. For every branch \(b\in2^\omega\), let \(R\) be the set of states occurring infinitely often along \(b\) in \(\rho\), and let \(a=t(b)\). We replace the labels of \(s\) and \(\rho\) on \(b\) by the pair \((c,q)\) fixed for \((R,a)\). Such a pair exists because the original \(s\) and \(\rho\) satisfy \(f(s(b))=t(b)\) and \((R,s(b),\rho(b))\in\Delta_\omega\). The resulting input tree still projects to \(t\), and the resulting run is accepting. It remains to verify the required topological complexity. For every \(R\subseteq Q\), the set of branches on which exactly the states from \(R\) occur infinitely often is a Boolean combination of \(F_\sigma\) sets, being defined by finitely many Büchi and co-Büchi conditions. Since \(t\in\fsigmatree\Sigma\), the set of branches carrying any particular input letter \(a\in\Sigma\) is also a Boolean combination of \(F_\sigma\) sets. Therefore, for every \(c\in\Gamma\) and \(q\in Q\), the sets of branches assigned the labels \(c\) and \(q\), respectively, are finite unions of intersections of Boolean combinations of \(F_\sigma\) sets, and are thus themselves Boolean combinations of \(F_\sigma\) sets. Hence the modified input tree belongs to \(\fsigmatree\Gamma\), and the modified run is a Borel tree. This proves \cref{item:fsigma}.
    \end{claimproof}
    This concludes the proof of \cref{thm:f-sigma}.
\end{proof}

We can recast the above theorem, or more precisely the slightly stronger statement of~\eqref{eq:two-semantics}, as follows. We can view "Borel \mso" as the first-order theory of the structure where the universe consists of Borel subsets of $2^{\leq \omega}$. Then~\eqref{eq:two-semantics} tells us that if we further restrict the universe to Boolean combinations of $\Fsigma$ sets, then we get an elementary substructure.
As we have remarked in the introduction, this type of situation does arise in the weaker setting of "Borel \mso" on $2^\omega$, and therefore it is plausible that it could also occur for $2^{\leq \omega}$.

\section{Prioritized Borel games}
\label{sec:prioritized-borel-games}
As we have explained in the previous section, the crucial point of our approach is finite-memory determinacy for acceptance games that arise from automata complementation. The rest of this paper is devoted to a study of these games, and evidence for the conjecture that they are determined with finite memory.

\subsection{Definition and finite-memory conjecture}
Before continuing, we clean up the definition of the games, by removing certain artifacts that arise from automata and which are not important to understand the games. The cleaned up version is presented below; one of its advantages is that it replaces the Muller condition on states by priorities on edges, as in parity conditions, which makes it more convenient to identify relevant special cases.

\begin{definition}
    [Prioritized Borel game] \AP A ""prioritized Borel game"" is defined in the same way as a "Muller-Borel game", except that the set of edge labels is \(\{1,\ldots,k\}\), for some \(k\in\mathbb N\), and the winning condition in \cref{item:muller-borel-condition} of \cref{def:muller-borel} is replaced by:
    \begin{enumerate}
    	\setcounter{enumi}{1} 
        \item \textbf{Prioritized Borel condition.} The winning condition is described by a sequence of Borel sets
        \begin{align*}
        W_1,\ldots,W_k \subseteq 2^\omega.
        \end{align*}
        The labels \(1,\ldots,k\) are called ""priorities"". For an infinite play, let \(i\) be the maximal priority that occurs infinitely often on its edges.
        Then Eve wins if the branch of the play belongs to $W_i$, and Adam wins otherwise.
    \end{enumerate}
\end{definition}

Note that the tree-like structure influences the information that is explicitly available for the players during a "play".
Namely, the sequence of produced nodes of the tree is encoded in the arena, while the sequence of states or "priorities" is not; players need memory to keep track of this latter information.

In the remainder of this paper we discuss the following conjecture.
\begin{conjecture}\label{conj:prioritized-memory}
    For every "prioritized Borel game", one of the players has a winning strategy that uses "finite memory".
    \AP Furthermore,
    an upper bound on the memory size can be computed based on the number of ""states"" (i.e., elements of $Q$) and number of "priorities".
\end{conjecture}

\AP We note that "prioritized Borel games" are "determined", that is, there is a ``""winner""'', as they are a special case of games with Borel winning conditions. The interesting part of the conjecture is therefore the one concerning memory.

It is easy to see that this conjecture is equivalent to \cref{conj:muller-borel}. In one direction, a Boolean combination of \buchi conditions can be translated into a prioritized condition using, e.g., a ``latest appearance record'' construction. In the other direction, a prioritized Borel game can be viewed as a Muller-Borel game by extending the state space so that each state additionally records the priority of the incoming edge. Whether a given priority occurs infinitely often can then be expressed by a \buchi condition on the extended states. In both constructions, a strategy for the new game can be implemented in the original game by additionally storing in memory the information encoded in the new states. Since the number of possible values of this information is computably bounded, the corresponding memory bounds remain computable.

\begin{remark}
    If every "prioritized Borel game" can be won with finite memory, then for a fixed number of "states" and priorities there is a uniform bound on the memory needed.
    That is, for all integers $n$ and $k$, there is a constant $c_{n,k}$ such that in any "prioritized Borel game" with $|Q|\leq n$ and using $k$ "priorities", one of the players has a winning strategy that uses at most $c_{n,k}$ memory states.

    Indeed, if this was not the case, there would be a sequence of games where the winner (that we can assume to be Eve by symmetry) requires unbounded memory.
    We could combine these games into a single one: Adam can stay in the leftmost branch and move into the $i$-th game at the $i$-th node. If he stays in the leftmost branch forever, he loses.
\end{remark}

\begin{example}[One state and priority]\label{ex:one-state-conjecture}
To get a feeling for the conjecture, let us discuss the special case where there is one state and one priority. In this case, the finite-memory assumption trivialises, so the above conjecture is just a restatement of Borel determinacy.
Indeed, in this case there is no difference between general strategies and "finite-memory strategies" (even "memoryless" strategies), since the current position determines the entire history.
\end{example}

Next, we show an example of a "prioritized Borel game" where the winner requires memory.

\begin{figure}[h]
            \centering
			\includegraphics[width=0.2\textwidth]{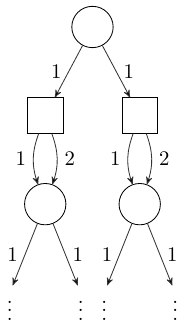}
		  \caption{
			The "prioritized Borel game" from \cref{ex:copying-game}. Eve nodes are represented as circles, and Adam nodes as squares.
		  }\label{fig:game-alternation}		
\end{figure}

\begin{example}[Copying game]\label{ex:copying-game}
    In \Cref{fig:game-alternation} we show a "prioritized Borel game" with only one state per tree node, and using two priorities.
    In this game, Adam chooses to produce a priority $1$ or $2$, and then Eve chooses to go left or right in the tree.
    We let the Eve-winning Borel subsets be $W_1 =$ branches that eventually only take the left child, and $W_2 =$ branches that contain a right child infinitely often.
    That is, Eve wins if she goes right infinitely often if and only if Adam produces $2$ infinitely often.

    Eve can win this game: each time that Adam takes a $2$-edge, she can go to the right child in the tree.
    However, she needs memory to implement this strategy.
    In fact, she cannot win with a memoryless strategy, as such a strategy determines one branch, and Adam can produce only priorities $1$ or $2$ depending on whether right children occur infinitely often in the branch.
\end{example}

\subsection{Upper bounds: Finite-memory results}
The rest of this paper is dedicated to a discussion of \cref{conj:prioritized-memory}.
In this section, we give some positive evidence, by proving some special cases of the conjecture. In \cref{sec:lower-bounds}, we exhibit some challenges in proving the conjecture, in particular we show that unbounded memory is necessary (\cref{prop:unbounded-memory}).
Some further special cases can be found in Manthe's PhD thesis~\cite[Chapter~3.4]{Manthe26Thesis}.

\subsubsection{One priority}\label{subsec:one-priority} We start discussing the case of "prioritized Borel games" with just one "priority".
Although this case is relatively simple, it illustrates well our general strategy for tackling \cref{conj:prioritized-memory}.
In this case, there is only one Borel winning set $W_1$ and the only condition for Eve to win is that the final branch lies in this set.

\begin{proposition}\label{prop:one-priority}
    In every "prioritized Borel game" with one priority, the "winner" has a "memoryless" strategy.
\end{proposition}

\begin{proof}
    Assume that Eve wins a given game with one priority using a strategy~$\sigma$.
    Let $W_\sigma\subseteq 2^\omega$ be the subset of branches that can be produced when Eve follows the strategy $\sigma$.

    Clearly $W_\sigma\subseteq W_1$, as the strategy is winning.
    Observe also that $W_\sigma$ is the set of branches that do not contain nodes from the set
    \[N = \{w\in 2^* \mid w \text{ is unattainable using } \sigma\}.\]
    Indeed, one inclusion is trivial, and for the other we use \konig lemma: if every node on a branch is attainable using $\sigma$, then there exists a single infinite play that is consistent with $\sigma$ and stays on that branch.

    We improve $\sigma$ to a memoryless strategy: in every position of Eve reached by many possible histories, we arbitrarily choose one of them, and follow the decision made by $\sigma$ after this history.
    Observe that the new strategy allows to reach only positions reachable by~$\sigma$; in particular no nodes from $N$ are visited, hence every branch consistent with the new strategy belongs to $W_\sigma\subseteq W_1$.
\end{proof}


Note that the set $W_\sigma$ in the proof above is topologically closed: it is the complement of an "open" set.
Thus, in a sense, we achieved our goal in this proof by replacing the set $W_1$, of arbitrary complexity in the Borel hierarchy, by a topologically simple one: a closed set, for which "memoryless" strategies suffice.

More generally, \cref{conj:prioritized-memory} holds for Borel subsets of ``low'' complexity.
If all winning sets $W_1,\dots, W_k$ are Boolean combinations of at most $m$ $\Fsigma$ sets, then the winner has a strategy that uses memory bounded by a constant that only depends on $k$ and $m$ (this follows from the classical finite-memory determinacy result of \buchi-Landweber~\cite{BL69Strategies} generalized to infinite arenas~\cite{GurevichHarrington82}).

We conjecture that whenever Eve wins a "prioritized Borel game", we can replace the winning sets $W_1,\dots, W_k$ by smaller subsets that are Boolean combinations of $\Fsigma$ sets of bounded complexity (depending on the number of states and priorities), in such a way that Eve still wins. This implies \cref{conj:prioritized-memory}. If Adam is the winner, we apply the conjecture to the dual game obtained by swapping the players and replacing each \(W_i\) by its complement. In fact, both conjectures are equivalent: given a finite-memory winning strategy, we can replace the winning sets by the sets that are consistent with the strategy (just as in the proof of \cref{prop:one-priority} above). These sets are of bounded topological complexity by \cite[Corollary 3.1]{borelmsodefbranches}. (Formally, the statement of the corollary is not sufficient, but we also need to obtain a bound on the complexity of the Boolean combination. This follows from its proof. See also~\cite[Theorem~3.5.3]{Manthe26Thesis}.)

\subsubsection{Two priorities}

We now prove \cref{conj:prioritized-memory} in the case of two priorities. \cref{ex:copying-game} uses two priorities, so we know that "memoryless" strategies do not suffice in this case.

\begin{proposition}\label{prop:two-priorities}
    In every "prioritized Borel game" with two priorities, the "winner" has a strategy using two states of memory.
\end{proposition}

\begin{proof}
        \AP
    Assume that Eve wins a given "prioritized Borel game" with two priorities using a strategy $\sigma$. We improve this strategy to a new one that uses only two states of memory. As in the previous proof, at each position reached by multiple histories consistent with~$\sigma$, we choose some of them (based only on the memory state), and we follow the decision of $\sigma$ made after this history.

    Compared to \cref{prop:one-priority}, we additionally have to ensure that we produce priority $2$ as often as $\sigma$ did: either finitely or infinitely often.
    We have two seemingly contradictory goals here: tend to produce priority $2$ when $\sigma$ did so, and tend not to produce priority~$2$ more often than $\sigma$ did.
    Intuitively, we use the memory bit to remember which goal we are currently realizing, and we switch between them from time to time.
    Namely, whenever our play produced priority $2$, we are temporarily happy with the first goal, and we switch to the second goal.
    But when all histories consistent with~$\sigma$ produced priority $2$, then we can (and should) also produce it, so we switch to the first goal.

    More formally, we identify some nodes (i.e., elements of $2^*$, not positions) as ""witnesses"". They are defined inductively as follows. The root is a witness. A non-root node $v \in 2^*$ is a witness if all plays consistent with $\sigma$ necessarily use "priority" $2$ between the previous witness node and $v$.

    The memory in the new strategy records whether "priority" $2$ has been produced since the last "witness", which takes two memory states: ``no'' and ``yes''.
    In particular, if we are in a witness node, then no priorities have been produced since the last witness, and hence the memory says ``no''.
    Then, for the $n$-th position of a play, based on this memory we define a  "history" $h_n$ as follows:
    \begin{itemize}
        \item Suppose that the memory says ``no'', that is, priority 2 has not been seen since the last witness.
              As $h_n$ we take any  "history" that ends in the current position, is consistent with $\sigma$, and has not seen priority 2 since the last witness.
        \item Suppose that the memory says ``yes'', that is, priority 2 has been seen since the last witness.
              \AP Define the ""profile@@two"" of a  "history" to be the sequence of priorities that were seen in the history.
	      Among all histories that reach the current position consistently with $\sigma$, as $h_n$ we choose one that has a maximal "profile@@two" lexicographically.
    \end{itemize}
    If it is Eve's move, she does whatever $\sigma$ would have done after $h_n$.
    If it is Adam's move, all his choices are valid continuations of $h_n$, consistent with $\sigma$.

    In order to prove that the new strategy is winning for Eve, consider some play $\pi$ that is consistent with this strategy, and consider three cases.

    (1) Suppose that $\pi$ visits infinitely many "witnesses". We conclude that Eve wins by showing that priority 2 is seen infinitely often along $\pi$, and that the branch of $\pi$ belongs to~$W_2$.

    To show that priority 2 occurs infinitely often in $\pi$, we prove that it occurs between every two consecutive witness nodes (we cannot simply use the definition of witness nodes here, since $\pi$ is not necessarily consistent with $\sigma$).
    Consider the position of $\pi$ just before the second witness.
    If $\pi$ did not produce priority $2$ yet, then also the  "history" $h_n$ defined in this position did not produce priority~2 since the last witness.
    The move of $\pi$ from this position extends~$h_n$ in a way consistent with $\sigma$, it thus produces priority $2$ by the definition of a witness.

    To show that the branch belongs to $W_2$,
    note that there are arbitrarily long histories (namely $h_n$) with the property: the  "history" is consistent with $\sigma$ and stays on the same branch as $\pi$.
    By \konig lemma, some infinite play also has this property.
    Since this play is consistent with $\sigma$, the definition of witnesses tells us that it sees priority 2 between every two witnesses.
    Thus, the branch must belong to $W_2$, since $\sigma$ is winning for Eve.

    (2) Suppose now that $\pi$ visits finitely many witnesses, and priority 2 is seen finitely often.
    Let $w \in 2^*$ be the last witness in the play.
    Because there are no more witnesses after $w$ in the branch, we can use the definition of witnesses to conclude that there are arbitrarily long histories with the property:
    the  "history" is consistent with $\sigma$, stays on the same branch as $\pi$, and does not see priority 2 after node $w$.
    Again by \konig lemma, there is some infinite play with this property.
    Since this play is winning for Eve, it follows that the branch is in $W_1$, and hence $\pi$ is also winning for Eve.

    (3) The last remaining case is when the play visits finitely many witnesses, but sees priority 2 infinitely often.
    In particular, priority~$2$ occurs after the last witness, and after this occurrence, each chosen  "history" $h_n$ is defined by maximising "profiles@@two".
    The  "history" $h_n$ extended by the next move of $\pi$ is one of the histories taken into account while defining $h_{n+1}$; we may thus only switch to a  "history" with a larger profile.
    This monotonicity ensures that the profiles of $h_n$ converge to some sequence $\alpha\in\{1,2\}^\omega$.
    Moreover, whenever $\pi$ contains priority $2$, then the profile of $h_{n+1}$ either contains $2$ on the last position, or is strictly greater than $h_n$ on earlier positions.
    This strict improvement occurs infinitely often, implying that the limit "profile@@two" $\alpha$ has infinitely many occurrences of priority~$2$.

    Now, the convergence means that there are arbitrarily long histories (appropriate prefixes of $h_n$) with the property:
    the  "history" is consistent with $\sigma$, stays on the same branch as $\pi$, and its profile is a prefix of $\alpha$.
    Another application of \konig lemma gives us an infinite play with this property.
    Since this play is winning for Eve, it follows that the branch is in $W_2$, and thus $\pi$ is winning for Eve as well.
\end{proof}

\begin{remark}\label{remark:topological-simplification}
    We discuss how this proof relates to the topological simplification approach described in \cref{subsec:one-priority}. For simplicity, we restrict to games with a single state. As in the proof of \cref{prop:one-priority}, let \(W_\sigma\) be the set of branches that can be produced when Eve follows \(\sigma\). Recall that \(W_\sigma\) is closed.

    Consider first priority \(1\). Let \(W'_1\) be the set of branches in \(W_\sigma\) that contain only finitely many "witnesses". By the argument from Case (2) above, \(W'_1\subseteq W_1\). Moreover, any play produced by the two-memory-state strategy in which priority \(2\) occurs only finitely often follows a branch in \(W'_1\), by Cases (1) and (2) above.

    We now define a corresponding subset \(W'_2\) of \(W_2\). At every node reachable by a history consistent with \(\sigma\), consider the lexicographically maximal "profile@@two" of such a history. We view each finite profile as an infinite sequence by padding it with ones. Along a branch, we say that the maximal profile increases or decreases whenever these padded profiles increase or decrease lexicographically from a node to its child. Let \(W'_2\) consist of those branches in \(W_\sigma\) for which either (i) there are infinitely many "witnesses", or (ii) the maximal profile decreases only finitely often and increases infinitely often.

    We have \(W'_2\subseteq W_2\). For (i), this follows from the argument in Case (1) above. For (ii), after the last decrease the maximal profiles form a nondecreasing sequence, and infinitely many strict increases imply that their limit contains infinitely many occurrences of priority~\(2\). As in Case (3), \konig lemma yields a play consistent with \(\sigma\), on the same branch, whose profile is this limit. Hence the branch belongs to \(W_2\).

    It remains to check that whenever a play produced by the two-memory-state strategy sees priority \(2\) infinitely often, its branch belongs to \(W'_2\). If there are infinitely many witnesses, Case (1) shows that the branch belongs to \(W_\sigma\), and hence it satisfies (i). Otherwise, after the last witness and the next occurrence of priority \(2\), the strategy always chooses histories with lexicographically maximal profiles. From then on the maximal profile never decreases, and every occurrence of priority \(2\) makes it increase. Thus the branch satisfies (ii); the argument of Case (3) also shows that it belongs to \(W_\sigma\). Together with the observation about \(W'_1\) above, this shows that the strategy constructed above remains winning after replacing \(W_1,W_2\) by their subsets \(W'_1,W'_2\).

    Finally, being a witness, an increase, or a decrease gives a fixed coloring of the nodes or edges of the tree. Having only finitely many occurrences of a given color is an \(F_\sigma\) condition, while having infinitely many is a \(G_\delta\) condition. Since \(W_\sigma\) is closed, \(W'_1\) is \(F_\sigma\), and \(W'_2\) is a Boolean combination of \(F_\sigma\) sets of bounded complexity.
\end{remark}

In the next section, we show that in games with $3$ priorities there is no constant bound on the size of the memory required, independent of the number of states (see \cref{prop:unbounded-memory}).
We have not succeeded in proving \cref{conj:prioritized-memory} for games with $3$ priorities.

\subsubsection{Nested branch conditions}

\AP We say that a sequence of subsets $W_1,\dots, W_k \subseteq 2^\omega$ is ""alternatingly nested"" if for every $1\leq p \leq k$,
\[W_p \subseteq \bigcap_{i<p} W_i\;\;\text{for odd $p$,\quad and}\quad W_p \supseteq \bigcup_{i<p} W_i\;\;\text{for even $p$.}\]

Note that the usual parity condition can be seen as a special case of an alternatingly nested condition: we take $W_p=\emptyset$ (i.e., Eve loses) for odd $p$ and $W_p=2^\omega$ (i.e., Eve wins) for even $p$.
The next proposition generalizes memoryless determinacy of parity games to prioritized Borel games with winning conditions of this form.

In the classical setting, Muller games can be reduced to parity games, using the latest appearance record~\cite{GurevichHarrington82}.
In the setting of this paper, we are not aware of any way of ensuring that the winning condition in a prioritized Borel game becomes alternatingly nested.
In particular, as already explained, the latest appearance record applied to "Muller-Borel games" gives us prioritized Borel games, but the resulting winning condition need not be "alternatingly nested".


\begin{proposition}\label{prop:nested}
    If the Borel sets of a "prioritized Borel game" are "alternatingly nested", then one of the players can win using a "memoryless strategy".
\end{proposition}

\begin{proof}
    It is enough to consider the case where Eve wins; the case of Adam is symmetric, after complementing the sets \(W_i\) and interchanging the roles of odd and even priorities. Let \(\sigma\) be a winning strategy for Eve. Once again, we improve it to a memoryless strategy by choosing one  "history" leading to a position, and performing the decision $\sigma$ made after this history.

    While proving \cref{prop:two-priorities}, it was neither better to have infinitely many nor finitely many occurrences of priority $2$; we thus had to use a memory state to alternate between two modes.
    For alternatingly nested conditions, it is always good to produce a priority $p$ infinitely often if $p$ is even, and finitely often if $p$ is odd.
    Thus, we should tend to produce even priorities when $\sigma$ can do so, and tend not to produce odd priorities more often than $\sigma$ did.
    To this end, we follow the worst possible history, defined next.

    First, we generalize the notion of a "profile@@two" from the previous proof to the case of more priorities.
    \AP Namely, for a priority $p$, the ""$p$-profile"" of a  "history"  $h$ is the sequence in $\{0,1\}^\omega$ having $1$ at indices for which priority $p$ was produced by $h$ (we pad the sequence with zeroes after the end of the history).
    To compare two histories $h_1$, $h_2$, we find the largest $p$ for which their "$p$-profiles" $\alpha_1, \alpha_2$ differ; then $h_1$ is \emph{worse} than $h_2$ if
    \begin{itemize}
    \item $p$ is odd, and $\alpha_1$ is lexicographically larger than $\alpha_2$ (intuitively: $h_1$ produces priority $p$ more often), or
    \item $p$ is even, and the last $1$ in $\alpha_1$ is earlier than in $\alpha_2$ (intuitively: $h_1$ produces priority $p$ less often), or
    \item $p$ is even, the last $1$ in $\alpha_1$ and $\alpha_2$ is at the same position, and $\alpha_1$ is lexicographically larger than $\alpha_2$ (just to break ties).
    \end{itemize}
    Then, among all histories that reach the current position consistently with $\sigma$, as $h_n$ we choose the worst possible  "history" (choosing arbitrarily among histories having the same "$p$-profiles" for all $p$).
    If it is Eve's move, she does whatever $\sigma$ would have done after $h_n$.

    In order to prove that the new strategy is winning for Eve, consider some play $\pi$ that is consistent with this strategy.
    For every $p\in\{0,1,\dots,k\}$ (by induction, starting from the highest priorities), we prove that either
    \begin{enumerate}[(1)]
    \item for all $j>p$, priority $j$ is seen finitely often in $\pi$ and, from some moment on, all histories $h_n$ have the same $j$-profile, or
    \item $\pi$ is winning for Eve.
    \end{enumerate}
    This is enough, since (1) cannot hold for $p=0$.

    It remains to prove the above. In the base case of $p=k$ this is trivial, because there are no larger priorities. For the induction step, fix some $p\in\{1,\dots,k\}$ for which the induction hypothesis already holds; we prove it for $p-1$. If (2) holds then there is nothing to prove. Otherwise, by (1), we know that priorities greater than $p$ are seen only finitely often in $\pi$ and, from some moment on, all histories $h_n$ have the same $j$-profile $\alpha_j$, for every $j>p$.
    Note that $\alpha_j$, being a profile of finite histories, may have $1$ only at finitely many positions.

    Suppose first that $p$ is even and priority $p$ occurs infinitely often in $\pi$.
    To show that the branch belongs to $W_p$,
    note that there are arbitrarily long histories (namely $h_n$) with the property: the  "history" is consistent with $\sigma$, stays on the same branch as $\pi$, and for every $j>p$ its $j$-profile is $\alpha_j$.
    By \konig lemma, some infinite play also has this property.
    It is a play won by Eve, where priorities greater than $p$ occur finitely often, which implies that the considered branch belongs to $W_i$ for some $i\leq p$.
    We conclude that Eve wins on $\pi$, thanks to the inclusion $W_i\subseteq W_p$.

    Next, suppose that $p$ is even, but priority $p$ occurs finitely often in $\pi$.
    Restrict our attention to positions after the moment where all $j$-profiles of $h_n$ for $j>p$ have stabilized, and after all occurrences in \(\pi\) of priorities greater than or equal to $p$.
    We observe that the "history" $h_n$ extended by the next move of $\pi$ is one of the "histories" taken into account while defining $h_{n+1}$.
    Since this move does not produce priority $p$, the history $h_{n+1}$ has either the same $p$-profile as $h_n$, or a worse one.
    The order used to define a worse history for even $p$ is well-founded, meaning that we can change to a worse "$p$-profile" only a finite number of times.
    Afterwards, the $p$-profile of successive histories $h_n$ stays the same; we thus obtain (1).

    We now switch to the case of odd $p$. Restrict our attention to positions after the moment where all \(j\)-profiles of \(h_n\) for \(j>p\) have stabilized, and after all occurrences in \(\pi\) of priorities greater than \(p\). We then have the same property as above: the history $h_{n+1}$ has either the same $p$-profile as $h_n$, or a worse one.
    It follows that the $p$-profiles converge monotonically to some limit sequence $\alpha_p$.
    Moreover, whenever priority $p$ occurs in $\pi$, the $p$\nobreakdash-profile of $h_{n+1}$ is strictly worse than $h_n$ (due to the history obtained by appending the move of $\pi$ producing priority $p$ to $h_n$).

    If the limit profile $\alpha_p$ has $1$ only on finitely many positions, then such a strict worsening could happen only finitely many times:
    the prefix of $\alpha_p$ containing all the ones becomes equal to prefixes of $p$-profiles of histories from some moment on, and the remaining suffix of zeroes is minimal lexicographically.
    This means that priority $p$ is seen finitely often in $\pi$ and, from some moment on, the $p$-profile of all further histories $h_n$ is $\alpha_p$; we obtain (1).

    The opposite case is that the limit profile $\alpha_p$ contains infinitely many ones.
    Then, the convergence gives us arbitrarily long histories (appropriate prefixes of $h_n$) with the property:
    the  "history" is consistent with $\sigma$, stays on the same branch as $\pi$, and for every $j\geq p$ its $j$-profile agrees with $\alpha_j$ up to the length of the history. By \konig lemma, there is also an infinite play consistent with \(\sigma\) and staying on the same branch as \(\pi\) whose \(j\)-profile is \(\alpha_j\) for every \(j\ge p\).
    It is a play won by Eve, where priorities greater than $p$ occur finitely often, but priority $p$ occurs infinitely often; in consequence the branch belongs to $W_p$.
    The largest priority occurring infinitely often on $\pi$ is some $i\leq p$; we conclude that Eve wins on $\pi$, thanks to the inclusion $W_p\subseteq W_i$.
\end{proof}

\subsubsection{One state}

The last upper bound we present, orthogonal to the previous ones, is a proof of \cref{conj:prioritized-memory} for the case of games with one state per tree node, that is, $|Q|=1$.

In games with one state, we identify positions of the game with tree nodes.
Note that, as "priorities"
appear on edges, despite the single-state restriction, a player may have up to $k$ non-equivalent moves going to the same target.
For this reason, a node in the game may admit multiple histories, leading to it from the root; each of these histories, leading to a~fixed target node, is completely determined by the sequence of "priorities" on it.

In the next section, we observe that already in games with only one state, Eve requires memory that depends on the number of "priorities" of the game (see \cref{prop:k-copying-game}).
The lower bound we provide there matches the upper bound we give in the next proposition.

\begin{proposition}\label{prop:single-state}
    In every "prioritized Borel game" with one state and $k$ "priorities", the "winner" has a strategy using at most $k$ states of memory.
\end{proposition}

Eve may face two kinds of decisions in her nodes: where to go (left or right child, if available), and which "priority" to pick among the ones available. (In the games in \Cref{fig:game-alternation,fig:game-k-copying}, only choices of the first kind appear.)
\AP By adding some intermediate nodes, we can isolate these decisions and assume that each Eve node is of one of the following forms:
\begin{description}
    \item[""Branching nodes"".] The node has a left and a right successor; in both directions the only "priority" is~$1$.
    \item[""Pick-priority nodes"".] The node has a single successor in the game, say the left child.
\end{description}
Adam nodes are unconstrained.

We extend the technique and notion of "witness" from the proof of the two-"priority" case: we label the tree with information about histories compatible with a fixed winning strategy~$\sigma$. Then, a strategy using finite memory has access to the information given by this labelling.
An important new insight is the use of "coding nodes".
Intuitively, we use some "branching nodes" to encode information about the past of the play---namely, a recent occurrence of a "priority" $i$.
Specifically, if Eve has the opportunity to go to an "$i$-coding node" $w$, she will do so precisely when the memory state tells her that ``"priority" $i$ has occurred recently''.
This allows us to redirect plays in which "priority" $i$ occurred in the recent past to the subtree of $w$, and in this way isolate them from other plays.

\begin{proof}[Proof of \Cref{prop:single-state}]
Consider a "game@@Borel" with a single state and $k$ "priorities", in which Eve wins.
We assume that nodes of the "game@@Borel" are of exactly one of the following three types: Adam nodes, "branching nodes" (which belong to Eve), or "pick-priority nodes" (which belong to Eve).

Fix a winning strategy $\sigma$ for Eve.
We define a strategy using $k$ memory states based on~$\sigma$.
Before that, we label the tree using two functions:
\[\witness:2^* \to \{\bot, 1,\dots, k \} \, , \quad \code:2^* \to \{\bot, 1,\dots, k \}. \]
\AP Nodes $w$ with $\witness(w)=\code(w)=\bot$ are called ""unattainable""; we define the new strategy so that it never reaches them.
In the remaining nodes $w$, which are called ""attainable"", both $\witness$ and $\code$ are numbers.
We say that an "attainable" node $w$ is an ""$i$-witness"" if $\witness(w) \geq i$ and that it is ""$i$-coding"" if $\code(w) \geq i$ (in particular, no node is an "$(k+1)$-witness" nor is "$(k+1)$-coding").
Moreover, we associate to each "attainable" node $w$ a history $h_w$ that leads to $w$ and is consistent with $\sigma$.

\AP For any node $w$ and any "priority" $i$, we write $\intro*\anc(w,i)$ for the most recent "$i$-witness" ancestor of~$w$ (possibly $w$ itself; in particular, $\anc(w,1) = w$ for all "attainable" nodes).
\AP We say that "priority" $i$ is ""recorded"" in $w$ if an "$i$-coding node" occurs on the path to $w$ strictly after $\anc(w,i)$.
\AP We say that a child $w'$ of $w$ is ""$i$-consistent"" if there is an extension of $h_{\anc(w,i)}$ that is consistent with $\sigma$, leads to $w'$, and produces "priority" $i$ as the highest "priority" after $\anc(w,i)$.

We define the labelling recursively.
First, in the root we put $\witness(\eps) = \code(\eps) = k$, and as $h_\eps$ we take the empty history. Next, assume that the labelling is already defined for a node $w$ and all its ancestors, and we want to define it for a child $w'$ of $w$.
If $w$ is "unattainable", we make $w'$ "unattainable" by setting $\witness(w') = \code(w') = \bot$.
Otherwise, as $\witness(w')$ and $\code(w')$ we take the least numbers satisfying the inequalities specified below:
\begin{enumerate}[(1)]
\renewcommand{\theenumi}{(\arabic{enumi})}\renewcommand{\labelenumi}{(\arabic{enumi})}
\item\label{labelling-item:recorded}
    For every "priority" $i$ that is "recorded" in $w$, if $w'$ is "$i$-consistent", we ensure that $\witness(w') \geq i$ and $\code(w') \geq i$.

\item\label{labelling-item:adam}
    Suppose that $w$ belongs to Adam.
    If there exists any "priority" $i$ for which $w'$ is "$i$-consistent", we choose the least such $i$ and ensure that $\witness(w') \geq i$ and $\code(w') \geq i$.

\item\label{labelling-item:pick-priority}
    Suppose that $w$ is a "pick-priority node".
    For every "priority" $i$ such that $w'$ is "$i$-consistent" and there is an edge $w\xrightarrow{i} w'$, we ensure that $\witness(w') \geq i$ and $\code(w') \geq i$.

\item\label{labelling-item:branching}
    Finally, suppose that $w$ is a "branching node".
    \AP 
    Note that $w$ has exactly one "$1$-consistent" child: the one picked by $\sigma$ after $h_w$.
    We denote the "$1$-consistent" child of $w$ by $\wpri$ and call it ""primary"" and the other child by $\wsec$ and call it ""secondary"".

    If $w'=\wpri$, we ensure that $\witness(w') \geq 1$ and $\code(w') \geq 1$.

    Suppose now that $w'=\wsec$ and that there exists a "priority" $i$ for which $w'$ is "$i$-consistent".
    We choose the least such $i$ and ensure $\witness(w') \geq i$ and $\code(w') \geq i$.
    We call $i$ the ""type"" of $w$.
    Let $r\in\{1,\dots,k\}$ be congruent modulo $k$ to the number of ancestors of $w$ that have the same "type" as~$w$ (note that not all nodes have "type", only some among "branching nodes").
    If the "priority" $r$ is not "recorded" in $w$, we additionally ensure that $\code(w') \geq r$.
\end{enumerate}
If none of the above conditions apply, we set $\witness(w') = \code(w') = \bot$.
Note however that if $\witness(w')=i\in\{1,\dots,k\}$ due to any of the above inequalities, then $w'$ is "$i$-consistent", that is, there is an extension of $h_{\anc(w,i)}$ that is consistent with $\sigma$, leads to $w'$, and produces "priority" $i$ as the highest "priority" after $\anc(w,i)$.
We take any such extension as~$h_{w'}$.

Observe that every "$i$-witness" is also "$i$-coding" (i.e., $\witness(w')\leq \code(w')$ for every "attainable" node $w'$).
Indeed, in the above definition, every inequality $\witness(w') \geq i$ is accompanied by the inequality $\code(w') \geq i$.

\begin{claim}\label{claim:history-extension-prev}
    If an "$i$-witness" $v$ is a proper ancestor of an "$i$-witness" $w'$ with $\witness(w') = i$, and no "$i$-witness" occurs strictly between them, then $h_{w'}$ extends $h_v$ and the highest "priority" in $h_{w'}$ after $v$ is $i$.
\end{claim}

\begin{claimproof}
    Let $w$ be the parent of $w'$; then $v=\anc(w,i)$.
    The claim follows directly from the definition of $h_{w'}$. 
\end{claimproof}

\begin{claim}\label{claim:history-extension-anc}
    For every "attainable" node $w$ and every "priority" $i$, the history $h_w$ extends $h_{\anc(w,i)}$ and its fragment after $\anc(w,i)$ produces only "priorities" smaller than $i$.
\end{claim}

\begin{claimproof}
    By induction on the level of $w$.
    Denote $j=\witness(w)$.
    If $j\geq i$, then $\anc(w,i)=w$ and the claim is trivial.
    Otherwise (i.e., when $j<i$), let $v$ be the most recent proper ancestor of $w$ that is an "$j$-witness".
    By \cref{claim:history-extension-prev}, $h_w$ extends $h_v$ and the highest "priority" in $h_w$ after~$v$ is $j$, hence all "priorities" produced there are smaller than $i$.
    By the induction hypothesis applied to $v$, the history $h_v$ (hence $h_w$ as well) extends $h_{\anc(v,i)}$ and its fragment between $\anc(v,i)$ and $v$ also produces only "priorities" smaller than $i$.
    We conclude observing that $\anc(w,i)=\anc(v,i)$, due to $j<i$.
\end{claimproof}

\begin{claim}\label{claim:child-consistent}
    Suppose that a history $h_w$ extended with an edge $w\xrightarrow{i}w'$ remains consistent with $\sigma$.
    Then the child $w'$ is "$i$-consistent".
\end{claim}

\begin{claimproof}
    The history $h_w$ together with the edge $w\xrightarrow{i}w'$ is consistent with $\sigma$ and, by \cref{claim:history-extension-anc}, it extends $h_{\anc(w,i)}$, and produces "priority"~$i$ as the highest "priority" after $\anc(w,i)$ ("priority"~$i$ occurs on the last edge, while earlier we have only "priorities" smaller than $i$); thus, $w'$ is "$i$-consistent".
\end{claimproof}

\begin{claim}\label{claim:pick-priority-ok}
    Let $w$ be an "attainable" "pick-priority node" and $w'$ its child.
    If $j$ is the minimal "priority" on an edge from $w$ to $w'$, then $w'$ is a "$j$-witness" (in particular, $w'$ is "attainable").
\end{claim}

\begin{claimproof}
    Let $i$ be the "priority" on the edge from $w$ to $w'$ played by $\sigma$ after $h_w$.
    \Cref{claim:child-consistent} implies that $w'$ is "$i$-consistent", so \cref{labelling-item:pick-priority} of the definition of the labelling enforces the inequality $\witness(w')\geq i$.
    Clearly $i\geq j$ (by minimality of $j$), so $w'$ is a "$j$-witness". 
\end{claimproof}

\begin{claim}\label{claim:fin-is-fin}
    If a branch contains only finitely many "$i$-witnesses", then it also contains only finitely many "$i$-coding nodes".
\end{claim}

\begin{claimproof}
    Consider a branch containing finitely many "$i$-witnesses".
    Let $j$ be the highest number for which the branch contains infinitely many "$j$-coding nodes"; we should prove that $j<i$.
    To the contrary, suppose that $j\geq i$.
    Then, let $w'$ be the second "$j$-coding node" after all the finitely many "$(j+1)$-coding nodes" and all the finitely many "$j$-witnesses" (due to $j\geq i$, every "$j$-witness" is also an "$i$-witness").
    In the definition of the labelling, the only inequality for $\code(w')$ that occurs without a corresponding inequality for $\witness(w')$ is $\code(w')\geq r$ in \cref{labelling-item:branching}; it was thus used with $r=j$.
    This inequality applies only when $j$ is not "recorded" in the parent $w$ of $w'$.
    However, we assumed that on the path to $w$ there already was some "$j$-coding node" after the last "$j$-witness", meaning that $j$ is "recorded" in $w$; a contradiction. 
\end{claimproof}

The new strategy that we define uses $k$ memory states, numbered $1,\dots,k$.
In the root, we start with memory state $1$.
Then, whenever the strategy has memory state $m$, and an edge $w\xrightarrow{i}w'$ is played between some "attainable" nodes $w, w'$ (the new strategy is defined so that only "attainable" nodes can be reached), we update the memory state as follows:
\begin{itemize}
\item if $i>m$ and $i$ is not "recorded" in $w$, we take $m'=i$; otherwise, we take $m'=m$;
\item if $w'$ is "$m'$-coding", as the new memory state we take $1$; otherwise, we take $m'$.
\end{itemize}

\begin{claim}\label{claim:memory-played}
    If the memory state in a node $w$ is $m\geq 2$, then "priority" $m$ was played between $\anc(w,m)$ and $w$, and $m$ is not "recorded" in $w$.
\end{claim}

\begin{claimproof}
    In the "$m$-coding node" $\anc(w,m)$, the memory state cannot be $m$ (it is either greater, or it is decreased to $1$).
    Later on, the memory state may become $m$ only when "priority" $m$ is played, just after a node in which priority $m$ is not "recorded".
    Any later occurrence of an $m$-coding node would reset the memory state to $1$, implying that priority $m$ remains not "recorded" until $w$.
\end{claimproof}

We are ready to define the new strategy.
While in node $w$ with memory state $m$, the choices of this strategy are as follows:
\begin{itemize}
\item
    Suppose that $w$ is a "pick-priority" node, and $i=\witness(w')$ for the unique child $w'$ of $w$ to which we can move.
    If there is an edge $w\xrightarrow{i} w'$, Eve takes it; otherwise, she takes the edge with the least possible "priority".
\item
    Suppose that $w$ is a "branching node" with "primary child" $\wpri$ and "secondary child" $\wsec$.
    Eve moves to $\wsec$ if
    \begin{itemize}
    \item $\code(\wsec) = m$ (then $\wsec$ ""demands to code""), or
    \item $\witness(\wsec)=\code(\wsec)=i$ for a "priority" $i$ that is "recorded" in $w$ (then $\wsec$ ""demands to witness"").
    \end{itemize}
    Otherwise, Eve moves to $\wpri$.
\end{itemize}

In the following, let $\pi$ be an infinite play consistent with the new strategy. We need to prove that $\pi$ is winning for Eve.

\begin{claim}\label{claim:no-high-coding}
    Let $i\in\{1,\dots,k\}$.
    If $\pi$ visits infinitely many "$i$-witnesses", then some "priority" $j\geq i$ is played infinitely often on $\pi$.
\end{claim}

\begin{claimproof}
    It is enough to prove the claim for the highest $i$ such that $\pi$ visits infinitely many "$i$-witnesses".
    \Cref{claim:fin-is-fin} implies that $\pi$ visits only finitely many "$(i+1)$-coding nodes".
    We prove that some "priority" $j\geq i$ is played in $\pi$ between every "$i$-witness" and the following "$i$-coding node" $w'$ (recall that every "$i$-witness" is "$i$-coding"), once we are below all "$(i+1)$-coding nodes"; this immediately gives the claim.

    Let $w$ be the parent of $w'$; then the considered "$i$-witness" is $\anc(w,i)$.
    Note also that "priority" $i$ is not "recorded" in $w$, since $w'$ is the first "$i$-coding node" in $\pi$ after $\anc(w,i)$.
    We now check the reason why $w'$ is "$i$-coding" (but not "$(i+1)$-coding"), which is some inequality $\code(w')\geq i$ in the definition of the labelling.

    It is not the inequality from \cref{labelling-item:recorded}, since that one only applies when $i$ is "recorded" in $w$.

    Suppose that $w$ belongs to Adam, and let $j$ be the "priority" played in $\pi$ between $w$ and $w'$.
    \Cref{claim:child-consistent} implies that $w'$ is "$j$-consistent".
    Recall that $i$ in \cref{labelling-item:adam} is chosen as the minimal number for which $w'$ is "$i$-consistent", hence the "priority" $j$ played in $\pi$ satisfies $j\geq i$.

    Next, suppose that $w$ is a "pick-priority node".
    The inequality $\code(w')\geq i$ in \cref{labelling-item:pick-priority} is enforced only when there exists an edge $w\xrightarrow{i}w'$; simultaneously, we obtain $\witness(w')=i$.
    In such a situation, the new strategy of Eve picks that edge, hence "priority" $i$ is played in $\pi$.

    Finally, suppose that $w$ is a "branching node" with "primary child" $\wpri$ and "secondary child"~$\wsec$.
    If $w'=\wpri$, then the inequality enforced in \cref{labelling-item:branching} is $\code(w')\geq 1$ (i.e., $i=1$), hence already the "priority" $1$ played just before $w'$ gives the thesis.
    Suppose thus that the node $w'$ chosen by Eve is $\wsec$.
    Note that $\wsec$ does not "demand to witness", since, as already observed, "priority" $i$ is not "recorded" in $w$.
    Thus $\wsec$ "demands to code", hence the memory state $m$ in $w$ equals $\code(\wsec)$, that is, $i$.
    Then, by \cref{claim:memory-played}, "priority" $i$ was played in $\pi$ between $\anc(w,i)$ and $w'$.
\end{claimproof}

Let $i_{\max}$ be the highest "priority" played infinitely often in the considered play $\pi$.
By \cref{claim:no-high-coding} we know that $\pi$ contains only finitely many "$(\imax+1)$-witnesses".

Suppose first that $\pi$ contains infinitely many "$\imax$-witnesses".
Let $w_1,w_2,\dots$ be the sequence of all "$\imax$-witnesses" in $\pi$ located after all "$(\imax+1)$-witnesses".
By \cref{claim:history-extension-prev}, each history $h_{w_{i+1}}$ extends the previous one, $h_{w_i}$, and the highest "priority" in $h_{w_{i+1}}$ between $w_i$ and $w_{i+1}$ is $\imax$.
In particular, there exists an infinite play that agrees with all the histories $h_{w_1},h_{w_2},\dots$.
This limit "play" $\pi'$ is consistent with $\sigma$, hence it is winning for Eve, and the highest "priority" played infinitely often in $\pi'$ is $\imax$.
Therefore, the play $\pi$, following the same branch and having the same highest "priority" played infinitely often, is also winning for Eve.

In the remaining part of the proof (including the next claims) we assume that $\pi$ visits only finitely many "$\imax$-witnesses".
In this case we have that $\imax\geq 2$, since every node is a "$1$-witness".

\begin{claim}\label{claim:infinitely-many-max-adam}
    Adam plays "priority" $\imax$ infinitely often.
\end{claim}

\begin{claimproof}
    We prove that Eve plays "priority" $\imax$ only finitely often; this gives the claim since altogether "priority" $\imax$ is played infinitely often.
    Recall that edges from "branching nodes" have "priority" $1$, hence "priority" $\imax\geq 2$ can be played only from "pick-priority nodes".
    The strategy may play "priority" $\imax$ from such a node $w$ to its child $w'$ for one of two reasons:
    \begin{itemize}
    \item $\imax=\witness(w')$, hence $w'$ is an "$\imax$-witness", or
    \item $\imax$ is the least "priority" on an edge from $w$ to $w'$; then again $w'$ is an "$\imax$-witness", by \cref{claim:pick-priority-ok}.
    \end{itemize} 
    We assumed that $\pi$ visits only finitely many "$\imax$-witnesses", so indeed this could happen only finitely often.
\end{claimproof}

\begin{claim}\label{claim:not-recorded}
    "Priority" $\imax$ is not "recorded" in any node in $\pi$ after the last "$\imax$-witness"; equivalently, after the last "$\imax$-witness" in $\pi$ there are no more "$\imax$-coding nodes".
\end{claim}

\begin{claimproof}
    To the contrary, suppose that $\imax$ is "recorded" in some node $v$ in $\pi$ after the last "$\imax$-witness".
    Consider a node $w$ after $v$ such that in $\pi$ Adam plays "priority" $\imax$ from $w$ to its child $w'$; it exists by \cref{claim:infinitely-many-max-adam}.
    \Cref{claim:child-consistent} implies that $w'$ is "$\imax$-consistent".
    There are no "$\imax$-witnesses" between $v$ and $w$, so $\imax$ is still "recorded" in $w$.
    Thus, \cref{labelling-item:recorded} of the definition of the labelling implies that $w'$ is an "$\imax$-witness", contradicting the assumption on $v$.
 \end{claimproof}

\begin{claim}\label{claim:memory-constant}
    During the play $\pi$, the memory state eventually remains constant and greater than or equal to $\imax$.
\end{claim}

\begin{claimproof}
    Consider a node $w$ after all the finitely many occurrences of "priorities" greater than~$\imax$, and after all the finitely many "$\imax$-coding nodes" (cf.\@ \cref{claim:fin-is-fin}), such that in $\pi$ Adam plays "priority" $\imax$ from $w$ to its child $w'$.
    Since $\imax$ is not "recorded" in $w$ (by \cref{claim:not-recorded}), the memory state is increased to $\imax$ in $w'$, unless it was already higher.
    Later on, the memory state can be increased only if we see a higher "priority", and can be decreased (to $1$) only if we visit an "$m$-coding node" for $m$ being the current memory state; none of these happen.
\end{claimproof}

Let $\minf\geq\imax$ be the ultimate value of the memory state, existing by \cref{claim:memory-constant}.

\begin{claim}\label{claim:finitely-many-consistent}
    The play $\pi$ visits only finitely many "branching nodes" such that their "secondary child" is "$i$-consistent" for some $i\leq\minf$.
\end{claim}

\begin{claimproof}
    We consider "branching nodes" of every "type" separately.
    First, $\pi$ may visit "branching nodes" without a "type", that is, such that their "secondary child" is not "$i$-consistent" for any $i$.
    They are simply not counted in the statement of the claim.
    Second, $\pi$ may visit "branching nodes" $w$ having "type" $t>\minf$.
    Recalling that the "type" of $w$ is the least "priority" $t$ for which the "secondary child" is "$t$-consistent", we again see that the "secondary child" of those nodes $w$ cannot be "$i$-consistent" for any $i\leq\minf$.

    Finally, consider "branching nodes" of some "type" $t\leq\minf$.
    We claim that $\pi$ visits only finitely many of them.
    Suppose otherwise.
    Then, $\pi$ visits also infinitely many nodes $w$ of "type" $t$ such that the number of ancestors of $w$ also having "type" $t$ is congruent to $\minf$ modulo~$k$.
    Consider any such $w$ that is located after all the finitely many "$\imax$-witnesses", and after the moment when the memory state becomes permanently $\minf$ (cf.\@ \cref{claim:memory-constant}); let $\wsec$ be the "secondary child" of $w$.
    By \cref{claim:memory-played}, $\minf$ is not "recorded" in $w$.
    Thus, \cref{labelling-item:branching} in the definition of our labelling enforces both inequalities $\code(\wsec)\geq t$ and $\code(\wsec)\geq\minf$ (only the latter counts, as $t\leq\minf$).
    We have two cases:
    \begin{itemize}
    \item
        If $\code(\wsec)=\minf$, then $\wsec$ "demands to code".
    \item
        Alternatively, we may have $\witness(\wsec)=\code(\wsec)=i$ for some "priority" $i>\minf$ that is recorded in $w$, due to inequalities from \cref{labelling-item:recorded} of the definition of the labelling.
        In this case, $\wsec$ "demands to witness".
    \end{itemize}
    In both cases, the new strategy chooses $\wsec$, hence $\pi$ goes there.
    However, $\wsec$ is "$\imax$-coding" (due to $\minf\geq\imax$ and, in the second case, $i>\minf$), contrarily to the assumption that we are after all "$\imax$-witnesses" in $\pi$, hence after all "$\imax$-coding nodes" (by \cref{claim:not-recorded}).
\end{claimproof}

Let $\wstar$ be a node in $\pi$ such that
\begin{itemize}
\item
    $\wstar$ is after all occurrences of "priorities" greater than $\imax$ (there are finitely many of them, by maximality of $\imax$),
\item
    $\wstar$ is after all "$\imax$-witnesses" in $\pi$ (by assumption there are finitely many of them),
\item
    starting from $\wstar$, the memory state remains $\minf$ forever (cf.\@ \Cref{claim:memory-constant}),
\item
    $\wstar$ is after all "branching nodes" such that their "secondary child" is "$i$-consistent" for some $i\leq\minf$ (there are finitely many of them, by \cref{claim:finitely-many-consistent}), and
\item
    $\wstar$ is an Adam node where he picks the "priority" $\imax$ in $\pi$ (there are infinitely many such nodes by \cref{claim:infinitely-many-max-adam}).
\end{itemize}

\begin{claim}\label{claim:similar-play}
    It is possible to extend the history $h_{\wstar}$ into an infinite play $\pi'$ consistent with $\sigma$ such that:
    \begin{enumerate}[(a)]
    \renewcommand{\theenumi}{(\alph{enumi})}\renewcommand{\labelenumi}{(\alph{enumi})}
    \item
        $\pi'$ follows the same branch as $\pi$,
    \item\label{claim-item:eve}
        starting from $\wstar$, Eve plays only "priorities" smaller than $\imax$ in $\pi'$, and
    \item\label{claim-item:adam}
        starting from $\wstar$, Adam plays the same moves in $\pi$ and in $\pi'$.
    \end{enumerate}
\end{claim}
\vspace{-\baselineskip}
\begin{claimproof}
    Note that there is at most one play $\pi'$ consistent with $\sigma$ and extending $h_{\wstar}$ where, starting from $\wstar$, Adam plays the same moves as in $\pi$.
    We need to show that, in this play, Eve plays only "priorities" smaller than $\imax$ in "pick-priority nodes", and that she stays in the same branch as $\pi$ in "branching nodes". We show this by induction.

    Assume that we have built a history $h'$ extending $h_{\wstar}$ as desired up to an Eve node $w$.
    The induction hypothesis implies that no "priorities" greater than $\imax$ appear in $h'$ after $\wstar$: Eve does not play them by \cref{claim-item:eve}, and Adam by \cref{claim-item:adam} and by the assumption that no such "priorities" occur in $\pi$ after $\wstar$.
    Moreover, the "priority" played by Adam just after $\wstar$, both in $\pi$ and in $\pi'$, is $\imax$ (note that $\wstar$ and $w$ belong to different players, so $w\neq\wstar$, hence this "priority" is indeed played).
    Thus, the highest "priority" in $h'$ after $\wstar$ is $\imax$.
    Recall that, for any $i\geq\imax$, there are no "$i$-witnesses" in $\pi$ after $\wstar$, so $\anc(w,i)=\anc(\wstar,i)$.
    Therefore, \cref{claim:history-extension-anc} implies that $h'$ extends $h_{\anc(w,i)}$ and its fragment after $\anc(w,i)$ produces no "priorities" greater than $i$.

    Suppose first that $w$ is a "pick-priority node", and assume by contradiction that the strategy~$\sigma$ after $h'$ chooses an edge of "priority" $i\geq\imax$ from $w$ to its child $w'$.
    The history $h'$ extended by this edge is consistent with $\sigma$, extends $h_{\anc(w,i)}$, and its fragment after $\anc(w,i)$ produces $i$ as the highest "priority".
    This implies that $w'$ is "$i$-consistent".
    In such a situation, \cref{labelling-item:pick-priority} of the definition of the labelling implies that $w'$ is an "$i$-witness", hence an "$\imax$-witness", contradicting the choice of $\wstar$.

    Next, suppose that $w$ is a "branching node", and that after $h'$ the strategy $\sigma$ goes to $w'$.
    By the definition of $\wstar$, the "secondary child" $\wsec$ of $w$ is not "$i$-consistent" for any $i\leq\imax$ (recall that $\imax\leq\minf$).
    The history $h'$ extended by the edge from $w$ to $w'$ (having "priority" $1$) is consistent with $\sigma$, extends $h_{\anc(w,\imax)}$, and its fragment after $\anc(w,\imax)$ produces $\imax$ as the highest "priority".
    This implies that $w'$ is "$\imax$-consistent"; thus $w'\neq\wsec$.
    As observed just after the definition of the labelling, if $\witness(\wsec)=i$, then $\wsec$ is "$i$-consistent", which is now possible only for $i>\imax$.
    However, we also know that $\pi$ (after $\wstar$) does not visit "$\imax$-witnesses".
    It follows that both $\pi$ and $\pi'$ go from $w$ to its "primary child".
\end{claimproof}

The play $\pi'$, constructed in \cref{claim:similar-play}, is consistent with $\sigma$, so it is winning for Eve.
Because the plays $\pi$ and $\pi'$ follow the same branch and for both plays the highest "priority" occurring infinitely often is $\imax$, we obtain that also $\pi$ is winning for Eve.
\end{proof}

\subsection{Lower bounds}\label{sec:lower-bounds}
This section presents some lower bounds on the memory required in "prioritized Borel games".
Namely, we show that the required memory size necessarily depends on both the number of "priorities" and the number of states of the game.
We hope that this will help guide future approaches to solving \cref{conj:prioritized-memory}.

For a cleaner presentation of the lower bounds, we use "prioritized Borel games" where the branching of the tree may be greater than $2$, but still finite and bounded. Formally, we let the set of positions be a subset of $C^* \times Q$, where $C$ is a finite set of child directions. These games can be simulated by "prioritized Borel games" over the binary tree, in a way that adds neither priorities nor states: it suffices to introduce $\lceil\log|C|\rceil-1$ intermediate levels between each two nodes of the tree of a given $C$-branching game.

\subsubsection{Many priorities, one state}
We generalize \cref{ex:copying-game} to games with $k$ priorities, for any $k$, and show that in these games Eve requires $k$ states of memory to win.

\begin{proposition}\label{prop:k-copying-game}
    For every $k$, there is a "prioritized Borel game" with a single state and $k$ priorities where Eve requires $k$ states of memory to win.
\end{proposition}

\begin{proof}
    Consider the ($k$-branching) "prioritized Borel game" using $k$ priorities depicted in \Cref{fig:game-k-copying}.
    As previously, we use circles to represent nodes owned by Eve, and squares to represent those owned by Adam.
    In this game, Adam picks a priority in $\{1,\dots,k\}$; then, Eve chooses to go to one of $k$ possible children. 
    Eve wins if the highest "priority" that occurs infinitely often coincides with the highest child direction that is taken infinitely often.
    Formally, $W_i$ is the set of branches where the $i$-th child is taken infinitely often, but for all $j>i$, the $j$-th child is taken only finitely often.
	
    \begin{figure}[h]
    \centering
            \centering
		    \includegraphics[width=0.3\textwidth]{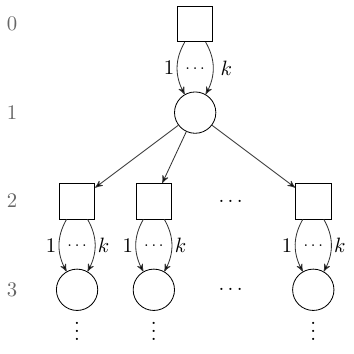}
		  \caption{
			The "prioritized Borel game" from \cref{prop:k-copying-game}.
		  }\label{fig:game-k-copying}
	  \end{figure}

    Eve can win this game by copying Adam's move (go to the $i$-th child if Adam picks priority $i$).
    This strategy uses $k$ memory states, which record the most recent "priority" played by Adam.

    We claim that this is optimal. Assume by contradiction that Eve has a winning strategy using $k-1$ memory states.
    Intuitively, in every step she fails to distinguish between two "priorities"; we will use this fact to build two plays on the same branch that produce a different maximal priority infinitely often.
    For each level $n$ of the tree, we define by recursion:
    \newcommand{\ipr}{i\rlap{$'$}} 
    \begin{itemize}
        \item a position $v_n$ of the game (controlled by Adam when $n$ is even, and by Eve when $n$ is odd),
        \item a memory state $m_n$, and
        \item two priorities, $i_n, \ipr_n$ (only for odd $n$).
    \end{itemize}
    We let $v_0$ be the root of the tree 
    and $m_0$ the initial memory state.
    Assume that these elements have been defined up to level $n$, for $n$ even ($v_n$ controlled by Adam).
    As there are only $k-1$ memory states, there are two priorities, $i_{n+1}$ and $\ipr_{n+1}$, such that the "memory update" from state $m_{n}$ when reading these two priorities goes to the same memory state: we let $m_{n+1}$ be this state.
    We let $v_{n+1}$ be the only successor of $v_n$ and $v_{n+2}$ be the child of $v_{n+1}$ picked by the strategy in the memory state $m_{n+1}$.
    Finally, let $m_{n+2}$ be the memory state obtained by updating $m_{n+1}$ over this last transition.

    This shows that, when Eve plays according to this strategy, Adam can decide to produce a play in the branch $v_0v_1v_2\dots$, while producing any sequence of "priorities" in $\{i_1,\ipr_1\}\{i_3,\ipr_3\}\{i_5,\ipr_5\}\dots$. Taking the sequences of the maximal and minimal elements of $\{i_n,\ipr_n\}$, respectively, we obtain two sequences with different largest "priority" occurring infinitely often, concluding the proof.
\end{proof}

\subsubsection{Three priorities, many states} In \cref{prop:two-priorities}, we showed that for games using two priorities, we can obtain a memory bound (two states) which is independent of the number of states of the game.
This is no longer true for three priorities.

\begin{proposition}\label{prop:unbounded-memory}
    For every $s$, there is a "prioritized Borel game" using three priorities where Eve requires $s+1$ memory states to win.
\end{proposition}

We define a "prioritized Borel game" $G_s$ where Eve requires $s+1$ memory states to win. We begin with an intuitive turn-based description of the game, which we later formalize as a "prioritized Borel game" with three priorities. In the game, the players construct a branch in a tree with branching $2\cdot s$. In order to avoid confusion with priorities, which are numbers, we use $C = \set{c_1,\ldots,c_{2s}}$ for the child directions in the tree. The players repeat infinitely often the following round consisting of three stages:
\begin{enumerate}
    \item Adam picks an even number $2i \in \set{2,4,\ldots,2s}$ and appends $c_{2i-1}$ to the current branch.
    \item Eve appends either $c_1$ or $c_{2i}$ to the current branch.
    \item Adam appends a finite sequence from $\set{c_1,\ldots,c_{2i}}^+$ to the current branch.
\end{enumerate}
In order to win, Eve needs to append $c_1$ infinitely often, and to ensure that the branch satisfies the parity condition, that is, the largest number appearing infinitely often is even.

Eve can win the game by alternating her choices ($c_1$ or $c_{2i}$) separately for each $i$. That is, for each $i \in \set{1,\ldots,s}$, she remembers which choice she made the last time $i$ was selected by Adam, and she uses the opposite choice. This strategy requires $2^s$ memory states. (Actually, Eve can win using $s+1$ memory states.) We will show below that, after the game is suitably formalised in our model, at least $s+1$ memory states are needed.

At every point, Eve has direct access to the finite prefix of the branch that has been produced up to that point. However, importantly, she does not have direct access to information about which players produced particular elements of the sequence, in particular when Eve appended $c_1$. She needs memory to recall this information. In a crucial way, Adam can add arbitrarily long finite sequences of nodes in his turns, which he can use to confuse Eve about the history of the game.

We now show that this game can be formalised using "prioritized Borel games". In our formalisation, we use three priorities $\set{1,2,3}$ to enforce that Adam plays finite sequences in each step, and that Eve appends $c_1$ infinitely often.

Here is the formal definition of the "prioritized Borel game". Each tree node has $2s+1$ states: $s+1$ of them controlled by Adam, $\hat q,q_1,\dots, q_s$; and $s$ of them controlled by Eve, $p_1,\dots,p_s$. The starting position is $(\eps,\hat q)$. When in state $\hat q$, Adam picks a number $i\in\{1,\dots,s\}$, emits priority $1$, descends in direction $c_{2i-1}$, and enters state $p_i$. When in a state $p_i$, Eve can choose to either emit priority $2$ and descend in direction $c_{2i}$, or to produce priority $3$ and descend in direction $c_1$; in both cases, she enters the state $q_i$. When in a state $q_i$, Adam emits priority $1$, descends in any direction from $\set{c_1,\ldots,c_{2i}}$, and enters either the state $q_i$ or the state $\hat q$ (in the first case, Adam intends to continue choosing children, while in the second case he finishes a round).

The winning condition is as follows. The set $W_1$ is full (if eventually only "priority" $1$ is produced, Eve wins). The set $W_2$ is empty (Adam always wins if $2$ is produced infinitely often and $3$ finitely often). If $3$ is produced infinitely often, Eve wins if and only if the highest child direction that occurs infinitely often is even. As we have explained earlier, Eve can win this game with finite memory. However, the amount of memory must grow with $s$, as the following lemma shows.

\begin{lemma}
    Eve cannot win the game $G_s$ using $s$ memory states.
\end{lemma}

\begin{proof}
    We prove the lower bound by induction on $s$. We use a slightly stronger induction hypothesis: Eve cannot win $G_s$ with a strategy that in every position with state $\hat q$ reaches at most $s$ different memory states.

    The base case of $s=1$ is resolved together with the induction step of $s\geq 2$. Assume, towards a contradiction, that in $G_s$ Eve has a winning strategy $\sigma$ with the property that, in every position with state $\hat q$, at most $s$ different memory states can be reached.

    We only consider plays starting as follows: from $(\eps,\hat q)$ Adam goes to position $(c_{2s-1},p_s)$, and then Eve plays according to her strategy $\sigma$. Let $w_0$ be the node reached after these two moves; it is either $c_{2s-1}c_1$ or $c_{2s-1}c_{2s}$.

    We say that a node $w$ ""allows priority $3$"" if for some history consistent with $\sigma$ that goes through position $(w,\hat q)$ and ends in position $(wc_{2s-1},p_s)$, Eve's strategy $\sigma$ from the latter position chooses to emit priority $3$ and descend in direction $c_1$.
    
    \begin{claim}\label{claim:allow-priority-3}
        There exists a node $w_1\in w_0C^*$ such that all its proper descendants "allow priority $3$".
    \end{claim}

    \begin{claimproof}
        Assume that this is not the case: every node $w_1$ in the subtree starting at $w_0$ has a proper descendant that does not "allow priority $3$". Then, we can build by induction a play consistent with $\sigma$ in which priority $3$ is never produced outside the root: In each round, first Adam descends in direction $c_{2s-1}$ entering state $p_s$, then Eve answers according to her strategy (already in the first round entering the subtree starting at $w_0$), and then Adam chooses a descendant of the current node that does not "allow priority $3$" and enters state $\hat q$ there. All nodes visited with state $\hat q$, except maybe the root, do not "allow priority $3$"; in other words, Eve in all her moves, except maybe the very first one, emits priority $2$. We have thus produced a play that is consistent with $\sigma$ and losing for Eve; a contradiction.
    \end{claimproof}

    Let $w_1$ be a node as in \cref{claim:allow-priority-3}. By taking it one level lower, we may assume that $w_1$ itself also "allows priority $3$", and that it is a proper descendant of $w_0$. We fix a function
    \[f\colon w_1C^*\to M\]
    that associates to each node $w$ in the subtree starting at $w_1$ some memory state reachable in $(w,\hat q)$ by a history consistent with $\sigma$ such that if from that position Adam goes to $(wc_{2s-1},p_s)$, then Eve's strategy $\sigma$ chooses to emit priority $3$ and descend in direction $c_1$. This function is well-defined since all $w\in w_1C^*$ "allow priority $3$".

    \begin{claim}\label{claim:induction-step}
        Every history consistent with $\sigma$ and ending in position $(w_2,\hat q)$ for some $w_2\in w_1C^*$ can be extended to a history consistent with $\sigma$ and ending in position $(w_3,\hat q)$ for some node $w_3\in w_2\{c_1,\dots,c_{2s-2}\}^*$ in such a way that the memory state in the target position $(w_3,\hat q)$ is $f(w_3)$.
    \end{claim}

    \begin{claimproof}
        Fix a history $h$ for which we want to prove the claim: a history consistent with $\sigma$ and ending in position $(w_2,\hat q)$ for some $w_2\in w_1C^*$. For $s=1$, we assumed that in every position with state $\hat q$ only one memory state can be reached by a history consistent with $\sigma$, so the memory state reached by $h$ in $(w_2,\hat q)$ is necessarily $f(w_2)$; we can thus take $w_3=w_2\in w_2\emptyset^*$.

        Suppose now that $s\geq 2$. In this case, if we consider the fragment of $G_s$ that starts in $(w_2,\hat q)$, and in which it is forbidden to use directions $c_{2s-1}$ and $c_{2s}$ (hence states $p_s$ and $q_s$), we obtain a game isomorphic to $G_{s-1}$. Moreover, $\sigma$ can be used to win (this copy of) $G_{s-1}$ if we identify histories in $G_{s-1}$ with histories in $G_s$ that start with $h$ and then continue as in $G_{s-1}$. Note that by removing directions $c_{2s-1}$, $c_{2s}$ and states $p_s$, $q_s$, we only eliminate some moves of Adam; there is no danger that the strategy of Eve will want to leave the considered subgame. Now, the induction hypothesis implies that every winning strategy of Eve in $G_{s-1}$ (hence, in particular, the one obtained from $\sigma$), in some position with state $\hat q$ reaches at least $s$ different memory states. Transferred back to $\sigma$, this means that for some node $w_3\in w_2\{c_1,\dots,c_{2s-2}\}^*$ there are extensions of $h$ consistent with $\sigma$ that reach at least $s$ different memory states in position $(w_3,\hat q)$. On the other hand, we assumed that in every position of $G_s$ with state $\hat q$ at most $s$ different memory states can be reached. It follows that some of the extensions reach in $(w_3,\hat q)$ the memory state $f(w_3)$, as needed for the claim.
    \end{claimproof}

    Consider now a play consistent with $\sigma$, built as follows. After the first two moves leading to $(w_0,q_s)$, Adam continues to $(w_1,\hat q)$. Then, we alternate forever between the following two phases. In the first phase, we use \cref{claim:induction-step} to reach a position $(w_3,\hat q)$ in memory state $f(w_3)$, appending only (zero or more) directions from $\set{c_1,\ldots,c_{2s-2}}$ to the current node address. In the second phase, Adam descends in direction $c_{2s-1}$ entering state $p_s$; then, by the definition of $f(w_3)$, according to $\sigma$ Eve answers by emitting priority $3$, descending in direction $c_1$, and entering state $q_s$; then, Adam descends in direction $c_1$ entering state $\hat q$, and we switch back to the first phase. Note that below $w_1$ the child direction $c_{2s}$ is never appended, while the child direction $c_{2s-1}$ is appended infinitely often. The play is thus losing for Eve; a contradiction.
 \end{proof}

\section{Generalizations}
Our upper bounds are proved by taking some winning strategy with possibly infinite memory,  and improving it so that it uses finite memory. In this sense, the Borel determinacy is used only to ensure that the former strategy exists. Therefore, one could consider a stronger variant of \cref{conj:muller-borel}, in which: (a) we no longer require the conditions $W \subseteq 2^\omega$ to be Borel, but we allow them to be arbitrary sets; and (b) we require that the game has a winner. This variant would be stronger than \cref{conj:muller-borel}, since (b) is automatic if we assume that the conditions $W \subseteq 2^\omega$ are Borel. The stronger variant of the conjecture  might be useful for sets beyond Borel sets, which still have determinacy results, such as analytic sets. (We do not use uncountable branching, which can be an obstruction to such generalizations.) Another related observation is that we do not depend on the full axiom of choice, and thus we could consider the full axiom of determinacy and the unrestricted monadic theory of $2^{\le\omega}$.

Yet another possible generalization involves the monadic theory of $2^{<\alpha}$ for various ordinals $\alpha$. Rabin's Tree Theorem is the  case of  $\alpha=\omega$, and this paper is about $\alpha = \omega+1$. Increasing $\alpha$ causes at least two kinds of additional difficulties. First, the behavior of automata for such a monadic theory along a single branch should be related to automata for words of length $\alpha$, which become more intricate for greater $\alpha$ \cite{ordinalsdecidable,neemanauto1,neemanauto2}. Second, our approaches require determinacy, which beyond $2^{<\omega\cdot2}$ requires large cardinals even for Borel sets \cite{longboreldet} and fails for games of length $\omega_1$ \cite[Theorem 7.3]{axdetorig}.

\bibliography{bibl.bib}

@phdthesis{Manthe26Thesis,
author = "Sven Manthe",
title = "Decidability in Borel Monadic Logic",
institution = "Rheinische Friedrich-Wilhelms-Universität Bonn",
year = "2026"
}

@inproceedings{Thomas95,
  author       = {Wolfgang Thomas},
  editor       = {Ernst W. Mayr and
                  Claude Puech},
  title        = {On the Synthesis of Strategies in Infinite Games},
  booktitle    = {{STACS} 95, 12th Annual Symposium on Theoretical Aspects of Computer
                  Science, Munich, Germany, March 2-4, 1995, Proceedings},
  series       = {Lecture Notes in Computer Science},
  volume       = {900},
  pages        = {1--13},
  publisher    = {Springer},
  year         = {1995},
  doi          = {10.1007/3-540-59042-0_57}
}

@misc{BookGamesOnGraphs2025,
      title={Games on Graphs: From Logic and Automata to Algorithms}, 
      author={Nathanaël Fijalkow and C. Aiswarya and Guy Avni and Nathalie Bertrand and Patricia Bouyer and Romain Brenguier and Arnaud Carayol and Antonio Casares and John Fearnley and Paul Gastin and Hugo Gimbert and Thomas A. Henzinger and Florian Horn and Rasmus Ibsen-Jensen and Nicolas Markey and Benjamin Monmege and Petr Novotný and Pierre Ohlmann and Mickael Randour and Ocan Sankur and Sylvain Schmitz and Olivier Serre and Mateusz Skomra and Nathalie Sznajder and Pierre Vandenhove},
      year={2026},
      eprint={2305.10546},
      archivePrefix={arXiv},
	  note = {Cambridge University Press},
    url = {https://www.cambridge.org/core/books/games-on-graphs/FBE35FACAC0A35949C996319E28DD0D7}
    }

@article{Landweber69OmegaAut,
	author       = {Lawrence H. Landweber},
	title        = {Decision Problems for omega-Automata},
	journal      = {{Math. Syst. Theory}},
	volume       = {3},
	number       = {4},
	pages        = {376--384},
	year         = {1969},
	doi          = {10.1007/BF01691063}
}

@article{Manthe24BorelMSO,
      title={The {B}orel monadic theory of order is decidable}, 
      author={Sven Manthe},
      year={2024},
      eprint={2410.00887},
      archivePrefix={arXiv},
      primaryClass={math.LO}
}

@article{Dzamonja25BorelMSO,
      title={{MSO} logic of the real order with the set quantifiers ranging over the {B}orel sets}, 
      author={Mirna Džamonja},
      year={2025},
      eprint={2512.23003},
      archivePrefix={arXiv},
      primaryClass={math.LO},
}

@article{BL69Strategies,
	ISSN = {00029947},
        doi = {10.2307/1994916},
	author = {J. Richard B{\"{u}}chi and Lawrence H. Landweber},
	journal = {{Transactions of the American Mathematical Society}},
	pages = {295--311},
	publisher = {American Mathematical Society},
	title = {Solving Sequential Conditions by Finite-State Strategies},
	volume = {138},
	year = {1969}
}

@incollection{Thomas97,
  address   = {Berlin},
  author    = {Thomas, Wolfgang},
  booktitle = {Handbook of formal languages, Vol.\ 3},
  pages     = {389--455},
  publisher = {Springer},
  title     = {Languages, automata, and logic},
  year      = 1997
}

@conference{GurevichHarrington82,
  author    = {Gurevich, Yuri and Harrington, Leo},
  booktitle = {Proc. ACM Symp. on Theory of Computing},
  pages     = {60--65},
  publisher = {Assoc. Comput. Mach.},
  title     = {Trees, automata and games},
  year      = 1982,
  doi = {10.1145/800070.802177}
}

@article{shelahord,
  author     = {Shelah, Saharon},
  title      = {The monadic theory of order},
  journal    = {Ann. of Math. (2)},
  fjournal   = {Annals of Mathematics. Second Series},
  volume     = {102},
  year       = {1975},
  number     = {3},
  pages      = {379--419},
  issn       = {0003-486X},
  mrclass    = {02G05 (02H10)},
  mrnumber   = {491120},
  mrreviewer = {Andreas\ Blass},
  doi        = {10.2307/1971037},
}

@article{monadicundecid,
  author     = {Gurevich, Yuri and Shelah, Saharon},
  title      = {Monadic theory of order and topology in {${\rm ZFC}$}},
  journal    = {Ann. Math. Logic},
  fjournal   = {Annals of Mathematical Logic},
  volume     = {23},
  year       = {1982},
  number     = {2-3},
  pages      = {179--198},
  issn       = {0003-4843},
  mrclass    = {03C85 (03C65 03F25 03F30 06A05 54A35)},
  mrnumber   = {701125},
  mrreviewer = {Alan\ Mekler},
  doi        = {10.1016/0003-4843(82)90004-3},
}

@incollection{monadicsurvey,
  editor     = {Barwise, J. and Feferman, S.},
  booktitle  = {Model-theoretic logics},
  series     = {Perspectives in Mathematical Logic},
  publisher  = {Springer-Verlag, New York},
  year       = {1985},
  isbn       = {0-387-90936-2},
  mrclass    = {03Cxx (03-02)},
  mrnumber   = {819531},
  mrreviewer = {P.\ \v{S}t\v{e}p\'{a}nek},
  author     = {Gurevich, Yuri},
  pages      = {479-506},
  title      = {Monadic Second-Order Theories}
}

@article{treeautomata,
  author     = {Rabin, Michael O.},
  title      = {Decidability of second-order theories and automata on infinite
                trees},
  journal    = {Trans. Amer. Math. Soc.},
  fjournal   = {Transactions of the American Mathematical Society},
  volume     = {141},
  year       = {1969},
  pages      = {1--35},
  issn       = {0002-9947,1088-6850},
  mrclass    = {02.32},
  mrnumber   = {246760},
  mrreviewer = {Walter\ Oberschelp},
  doi        = {10.2307/1995086},
}

@article{shelahquestions,
  author     = {Shelah, Saharon},
  title      = {On what {I} do not understand (and have something to say),
                model theory},
  journal    = {Math. Japon.},
  fjournal   = {Mathematica Japonica},
  volume     = {51},
  year       = {2000},
  number     = {2},
  pages      = {329--377},
  issn       = {0025-5513},
  mrclass    = {03Cxx (03-02)},
  mrnumber   = {1747306},
  mrreviewer = {O.\ V.\ Belegradek}
}

@article{borelmsodefbranches,
  author     = {Boja\'nczyk, Miko{\l}aj and Niwi\'nski, Damian and Rabinovich,
                Alexander and Radziwo\'nczyk-Syta, Adam and Skrzypczak,
                Micha{\l}},
  title      = {On the {B}orel complexity of {MSO} definable sets of branches},
  journal    = {Fund. Inform.},
  fjournal   = {Fundamenta Informaticae},
  volume     = {98},
  year       = {2010},
  number     = {4},
  pages      = {337--349},
  issn       = {0169-2968,1875-8681},
  mrclass    = {03D05 (03B15)},
  mrnumber   = {2650090},
  mrreviewer = {Markus\ Lohrey},
  doi        = {10.3233/fi-2010-231},
}

@article{Martin75,
  author  = {Martin, Donald A.},
  journal = {Ann. of Math. (2)},
  number  = 2,
  pages   = {363--371},
  title   = {Borel determinacy},
  volume  = 102,
  year    = 1975,
  doi={10.2307/1971035}
}

@article {ordinalsdecidable,
    AUTHOR = {B\"uchi, J. Richard and Zaiontz, Charles},
     TITLE = {Deterministic automata and the monadic theory of ordinals
              {$<\omega \sb{2}$}},
   JOURNAL = {Z. Math. Logik Grundlag. Math.},
  FJOURNAL = {Zeitschrift f\"ur Mathematische Logik und Grundlagen der
              Mathematik},
    VOLUME = {29},
      YEAR = {1983},
    NUMBER = {4},
     PAGES = {313--336},
      ISSN = {0044-3050},
   MRCLASS = {03B15 (03B25 03D05 03E10 68Q05)},
  MRNUMBER = {715254},
       DOI = {10.1002/malq.19830290503},
}

@article {neemanauto1,
    AUTHOR = {Neeman, Itay},
     TITLE = {Finite state automata and monadic definability of singular
              cardinals},
   JOURNAL = {J. Symbolic Logic},
  FJOURNAL = {The Journal of Symbolic Logic},
    VOLUME = {73},
      YEAR = {2008},
    NUMBER = {2},
     PAGES = {412--438},
      ISSN = {0022-4812,1943-5886},
   MRCLASS = {03E10 (03D05)},
  MRNUMBER = {2414457},
MRREVIEWER = {Arnold\ W.\ Miller},
       DOI = {10.2178/jsl/1208359052},
}

@incollection {neemanauto2,
    AUTHOR = {Neeman, Itay},
     TITLE = {Monadic definability of ordinals},
 BOOKTITLE = {Computational prospects of infinity. {P}art {II}. {P}resented
              talks},
    SERIES = {Lect. Notes Ser. Inst. Math. Sci. Natl. Univ. Singap.},
    VOLUME = {15},
     PAGES = {193--205},
 PUBLISHER = {World Sci. Publ., Hackensack, NJ},
      YEAR = {2008},
      ISBN = {978-981-279-654-7; 981-279-654-1},
   MRCLASS = {03E10 (03D05)},
  MRNUMBER = {2449465},
       DOI = {10.1142/9789812796554_0010},
}

@article {axdetorig,
    AUTHOR = {Mycielski, Jan},
     TITLE = {On the axiom of determinateness},
   JOURNAL = {Fund. Math.},
  FJOURNAL = {Polska Akademia Nauk. Fundamenta Mathematicae},
    VOLUME = {53},
      YEAR = {1963/64},
     PAGES = {205--224},
      ISSN = {0016-2736,1730-6329},
   MRCLASS = {02.60 (04.60)},
  MRNUMBER = {161787},
MRREVIEWER = {H.\ Jerome\ Keisler},
       DOI = {10.4064/fm-53-2-205-224},
}

@article {longboreldet,
    AUTHOR = {Aguilera, J. P.},
     TITLE = {Long {B}orel games},
   JOURNAL = {Israel J. Math.},
  FJOURNAL = {Israel Journal of Mathematics},
    VOLUME = {243},
      YEAR = {2021},
    NUMBER = {1},
     PAGES = {273--314},
      ISSN = {0021-2172,1565-8511},
   MRCLASS = {03E15 (03E55 91A44)},
  MRNUMBER = {4299148},
MRREVIEWER = {Xianghui\ Shi},
       DOI = {10.1007/s11856-021-2160-y},
}

\end{document}